\documentclass[oldversion]{aa}

\pdfoutput=1
\usepackage[tbtags,fleqn]{amsmath}
\usepackage{amsfonts}
\usepackage{graphicx}
\usepackage{natbib}
\usepackage{aalongtable}

\usepackage{framed,color,ocg}
\newcommand{\ToggleLayer}[2]{%
  \leavevmode
  \pdfstartlink user {
    /Subtype /Link
    /Border [0 0 0]%
    /A <<
      /S/JavaScript
      /JS (
         var aOCGs = this.getOCGs();
         var AllChecked = 1;
         var AllIndex = 0;
         for(var i=0; aOCGs && i<aOCGs.length;i++)
         {
         if(aOCGs[i].name == "#1") {aOCGs[i].state = !aOCGs[i].state;}
         if(!aOCGs[i].state && aOCGs[i].name != "all") {AllChecked = 0;}
         if(aOCGs[i].name == "all") {aOCGs[i].state = false; AllIndex=i;}
         }
         if(AllChecked == 1) {aOCGs[AllIndex].state = true;}
      )
    >>
  }#2%
  \pdfendlink
}

\newcommand{\diff}{\mathrm d}

\newcommand{\mincir}{\raise
  -2.truept\hbox{\rlap{\hbox{$\sim$}}\raise5.truept \hbox{$<$}\ }}
\newcommand{\magcir}{\raise
  -2.truept\hbox{\rlap{\hbox{$\sim$}}\raise5.truept \hbox{$>$}\ }}


\begin{document}

\title{2MASS wide field extinction maps: IV. The Orion, Mon R2,
  Rosette, and Canis Major star forming regions} 
\titlerunning{2MASS extinction maps: IV. Orion, Mon R2, Rosette, \&
  Canis Major} 
\author{Marco Lombardi\inst{1}, Jo\~ao Alves\inst{2}, and Charles
  J. Lada\inst{3}} \authorrunning{M. Lombardi \textit{et al}.}
\offprints{M. Lombardi} \mail{mlombard@eso.org} 
\institute{%
  University of Milan, Department of Physics, via Celoria 16, I-20133
  Milan, Italy \and 
  University of Vienna, T\"urkenschanzstrasse 17, 1180 Vienna, Austria
  \and Harvard-Smithsonian Center for
  Astrophysics, Mail Stop 72, 60 Garden Street, Cambridge, MA 02138}
\date{Received ***date***; Accepted ***date***}

\abstract{%
  We present a near-infrared extinction map of a large region
  (approximately $2\,200 \mbox{ deg}^2$) covering the Orion, the
  Monoceros R2, the Rosette, and the Canis Major molecular clouds.  We
  used robust and optimal methods to map the dust column density in
  the near-infrared (\textsc{Nicer} and \textsc{Nicest}) towards $\sim
  19$ million stars of the Two Micron All Sky Survey (2MASS) point
  source catalog.  Over the relevant regions of the field, we reached
  a 1-$\sigma$ error of $0.03 \mbox{ mag}$ in the $K$-band extinction
  with a resolution of $3 \mbox{ arcmin}$.  We measured the cloud
  distances by comparing the observed density of foreground stars with
  the prediction of galactic models, thus obtaining $d_\mathrm{Orion
    A} = (371 \pm 10) \mbox{ pc}$, $d_\mathrm{Orion B} = (398 \pm 12)
  \mbox{ pc}$, $d_\mathrm{Mon R2} = (905 \pm 37) \mbox{ pc}$,
  $d_\mathrm{Rosette} = (1330 \pm 48) \mbox{ pc}$, and $d_\mathrm{CMa}
  = (1150 \pm 64) \mbox{ pc}$, values that compare very well with
  independent estimates.  }

\maketitle

%

\defcitealias{2001A&A...377.1023L}{Paper~0}
\defcitealias{2006A&A...454..781L}{Paper~I}
\defcitealias{2008A&A...489..143L}{Paper~II}
\defcitealias{2010A&A...512A..67L}{Paper~III}

\section{Introduction}
\label{sec:introduction}

\begin{figure*}[!tbp]
  \centering
  \includegraphics[width=15cm]{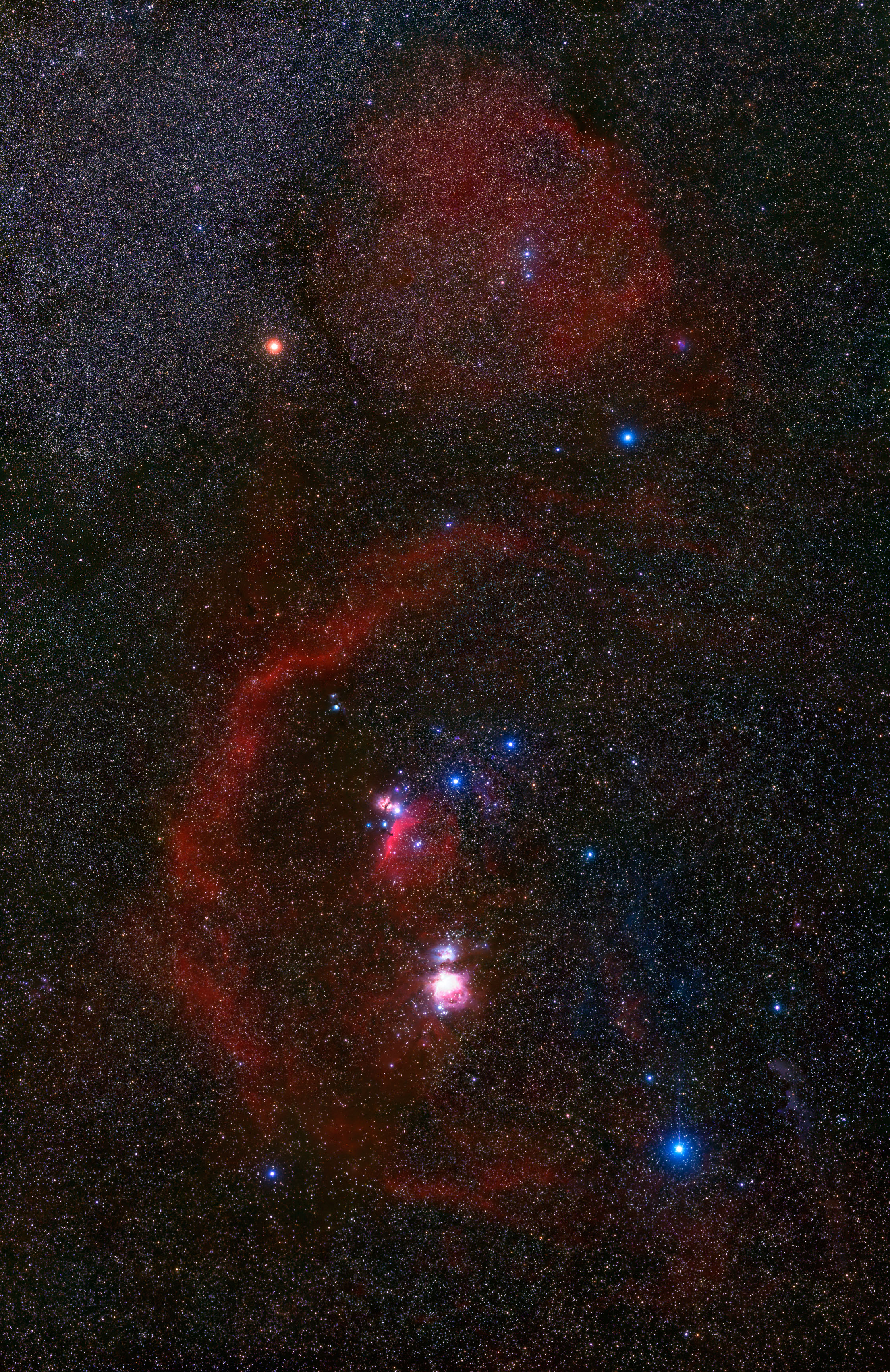}%
  \hspace{-15cm}%
  \begin{ocg}{clouds}{1}{1}%
    \includegraphics[width=15cm]{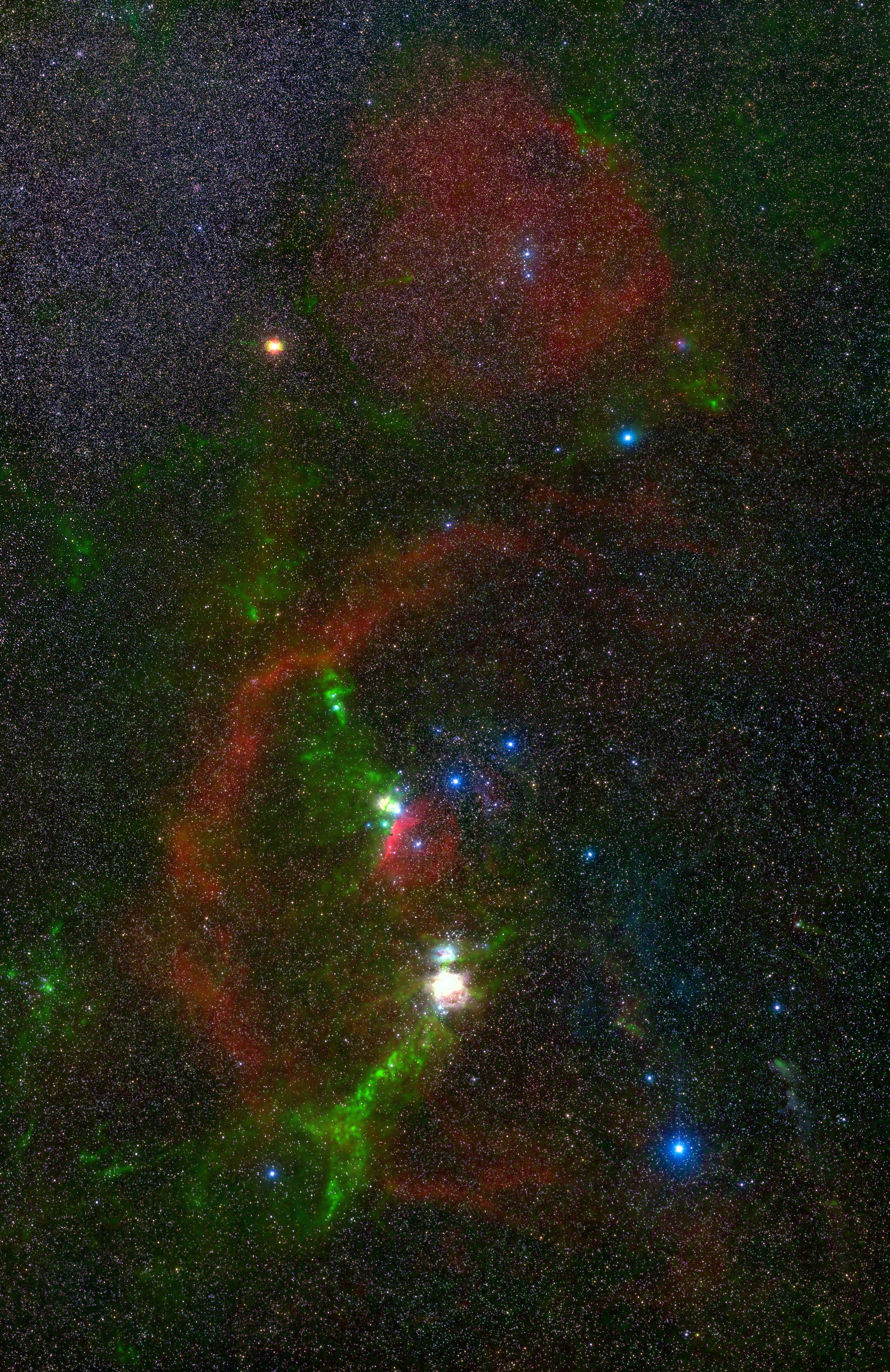}%
  \end{ocg}%
  \hspace{-15cm}%
  \begin{ocg}{labels}{2}{1}%
    \includegraphics[width=15cm]{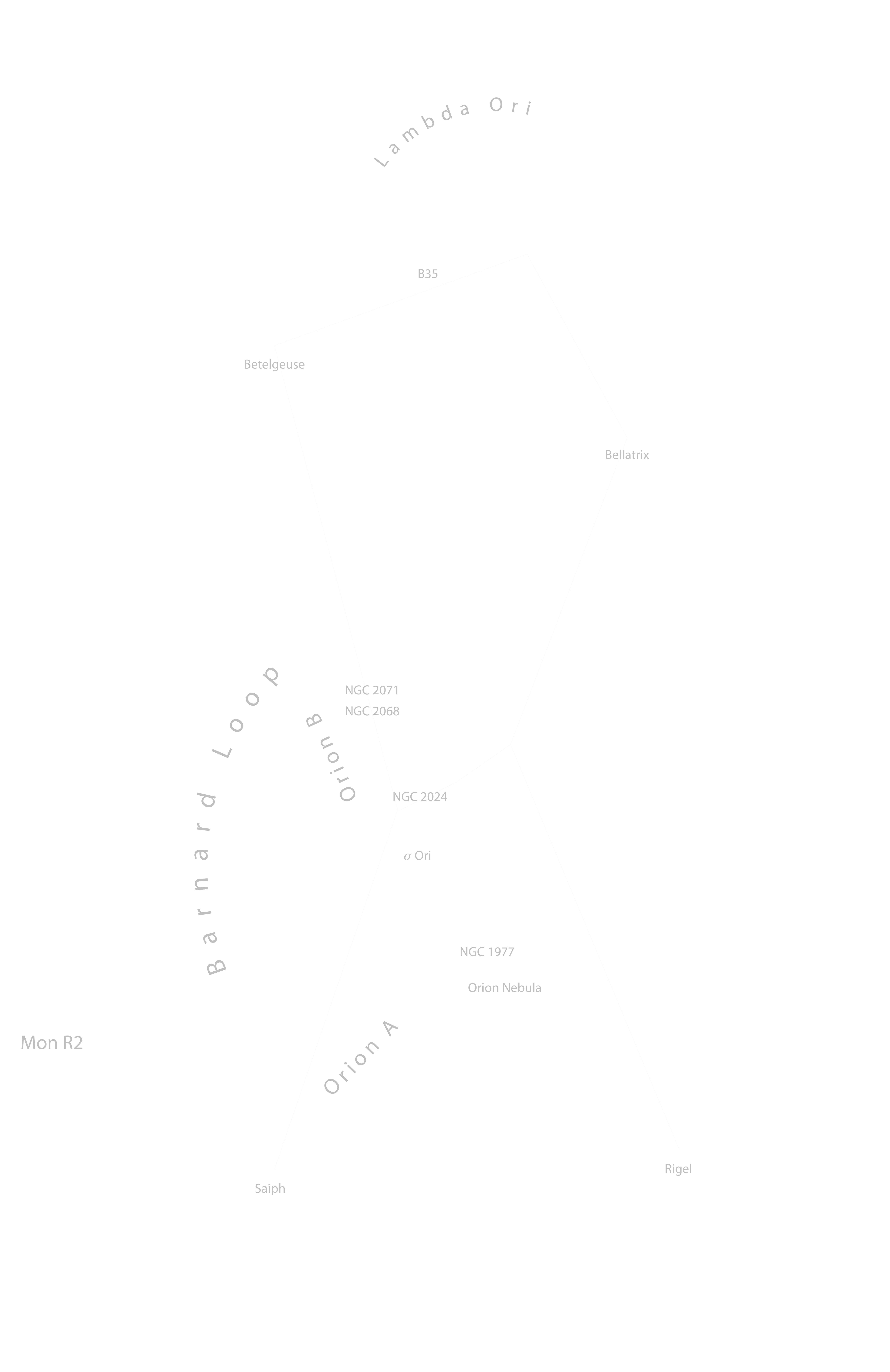}%
  \end{ocg}%
  \caption{An optical image of Orion (Wei-Hao Wang, IfA, University of
    Hawaii) with the extinction map presented in this paper
    superimposed in green.  The complementarity between the red
    \textsc{Hii} regions and the green $\mathrm{H}_2$ ones is evident.
    Note that the online figure is interactive:
    \ToggleLayer{clouds}{\protect\framebox{extinction}} and
    \ToggleLayer{labels}{\protect\framebox{labels}} can be toggled by
    clicking on the respective boxes when using
    Adobe$\textsuperscript{\textregistered}$
    Acrobat$\textsuperscript{\textregistered}$.}
  \label{fig:1}
\end{figure*}

In a series of papers, we have applied an optimized multi-band
technique dubbed Near-Infrared Color Excess Revisited (\textsc{Nicer}
\citealp{2001A&A...377.1023L}, hereafter Paper~0) to measure dust
extinction and investigate the structure of nearby molecular dark
clouds using the Two Micron All Sky Survey (2MASS;
\citealp{1994ExA.....3...65K}).  The main aim of our coordinated study
is to investigate the large-scale structure of these objects and to
clarify the link between the global physical properties of molecular
clouds and their ability to form stars.  Previously, we considered the
Pipe nebula (see \citealp{2006A&A...454..781L}, hereafter Paper~I),
the Ophiuchus and Lupus complexes (\citealp{2008A&A...489..143L},
hereafter Paper~II), and the Taurus, Perseus, and California complexes
(\citealp{2010A&A...512A..67L}, hereafter Paper~III).  We now present
an analysis of a large region covering more than $2\,200$ square
degrees, centered around Orion.  This region includes Orion~A and B,
$\lambda$ Orionis, Mon R2, Rosette, and Canis Major.  An overview of a
subset of the wide field extinction map presented in this paper,
superimposed on an optical image of the sky, is presented in
Fig.~\ref{fig:1}.

Near-infrared dust extinction measurement techniques present several
advantages with respect to other column density tracers.  As shown by
\citet{2009ApJ...692...91G}, observations of dust are a better column
density tracer than observations of molecular gas (CO), and
observations of dust extinction in particular provide more robust
measurements of column density than observations of dust emission,
mainly because of the dependence of the latter measurements on
uncertain knowledge of dust temperatures and emissivities.
Additionally, the sensitivity reached by near-infrared dust extinction
techniques, and by \textsc{Nicer} in particular, is essential to
investigate the low-density regions of molecular clouds (which are
often below the column density threshold required for the detection of
the CO molecule) and therefore to estimate the mass of the diffuse gas
that acts as a pressure boundary around the clumps.  Similar to
Paper~III, we use for some key analyses in this paper the improved
\textsc{Nicest} method \citep{2009A&A...493..735L}, designed to cope
better with the unresolved inhomogeneities present in the high-column
density regions of the maps.

Orion is probably the best studied molecular cloud in the sky.  The
complex comprises \textsc{Hii} and \textsc{Hi} regions superimposed on
colder, massive H$_2$ clouds with active formation of both low and
high mass stars.  The area studied here contains the Orion OB
association, which is split in several subgroups with ages from $2$ to
$12 \mbox{ Myr}$.  It is estimated that in the last $12 \mbox{ Myr}$
there have been 10 to 20 supernov\ae\ explosions
\citep{2008hsf1.book..459B} that shaped the gas in the region.  More
than a century ago, Barnard discovered a large arc of H$_\alpha$
emission around the eastern part of Orion (see Fig.~\ref{fig:1}).
More recently, Barnard's loop has been linked to Eridanus loop, and
the two have been identified as a superbubble that is expanding into
denser regions with a mean velocity of $10$--$20 \mbox{ km s}^{-1}$
\citep{1988ApJ...324..776M}.  Orion includes two giant molecular
molecular clouds, Orion~A and Orion~B, the spectacular $\lambda$
Orionis bubble, and a large number of smaller cometary clouds.  Both
Orion~A and B are observed in projection inside Barnard's loop and
most likely have been shaped by the supernova explosions, stellar
winds, and \textsc{Hii} regions generated by OB stars in the area;
for example, note the shell originating from the south-east
extremity of Orion~A, around the early B star $\kappa$ Orionis
(Saiph).  Orion~A and B host rich clusters of young ($\sim 2 \mbox{
  Myr}$) stars, the Orion Nebula Cluster (ONC), NGC~2024 (also known
as the Flame Nebula), NGC~2071, and NGC~2068.  Just southwest of
NGC~2024 is the famous Horsehead Nebula clearly visible as a
protrusion of the green extinction map in Fig.~\ref{fig:1}.

Around the ``head'' of Orion, the O8 star $\lambda$ Orionis, there is
a ring of dark clouds, $60 \mbox{ pc}$ diameter, known for almost a
century.  The region hosts several young stars (including 11 OB stars
very close to $\lambda$ Orionis), and their spatial and age
distributions show that originally star formation occurred in an
elongated giant molecular cloud \citep{2008hsf1.book..757M,
  1982ApJ...261..135D}.  Most likely, the explosion of a supernova
coupled with stellar winds and \textsc{Hii} regions destroyed the
dense central core, created an ionized bubble and a molecular shell
visible as a ring.  Star formation still continues in remnant dark
clouds distant from the original core.

The Monoceros R2 region is distinguished by a chain of reflection
nebul\ae\ that extend over $2^\circ$ on the sky.  The nomenclature
``Mon R2'' indicates the second association of reflection nebul\ae\ in
the constellation Monoceros \citet{1966AJ.....71..990V}.  The
region is also well known for its dark nebula, which is clearly
visible on top of the reflection nebul\ae\ and the field stars.
Clusters of newly formed stars are present in its core and in the GGD
12-15 region; smaller cores are also present in the field.  The cloud
is estimated to be at a distance significantly higher than the Orion
nebula, $830 \pm 50 \mbox{ pc}$ \citep{1976AJ.....81..840H}.

Rosette is known to amateur astronomers as one of the most spectacular
nebul\ae\ in the sky.  The region is characterized by an expanding
\textsc{Hii} region interacting with a giant molecular cloud.  The
\textsc{Hii} region has been photodissociated by NGC~2244, a cluster
containing more than 30 high-mass OB stars.  Judging from the age of
NGC~2244, the Rosette molecular cloud has been actively forming stars
for the last 2 to 3 Myr, and most likely will continue doing so, as
there is still sufficient molecular material placed in a heavily
stimulated environment \citep{2008hsf1.book..928R}.  The distance to
NGC~2244, and therefore to the molecular cloud, is still
controversial: recent estimates range between $1\,390 \mbox{ pc}$
\citep{2000A&A...358..553H} and $1\,670 \mbox{ pc}$
\citep{2002AJ....123..892P}.

The star-forming region in Canis Major is characterized by a
concentrated group of early type stars (forming the CMa OB1/R1
associations) and by an arc-shaped molecular cloud (probably produced
by a supernova explosion).  Similarly to Rosette, Canis Major presents
an interface between the \textsc{Hii} region and the neutral gas which
shows up as a north-south oriented ridge.  Recent distance
determinations of Canis Major seem to converge around a distance of
$1\,000 \mbox{ pc}$ \citep{2008hsf2.book....1G}.

This paper is organized as follows.  In
Sect.~\ref{sec:nicer-absorpt-map} we briefly describe the technique
used to map the dust and we present the main results obtained.
Section~\ref{sec:statistical-analysis} is devoted to an in-depth
statistical analysis, and includes a discussion of the measured
reddening laws for the various clouds, the measurements of the
distances using foreground stars, the log-normality of the column
density distributions, and the effects of small-scale inhomogeneities.
The mass estimates for the clouds are presented in
Sect.~\ref{sec:mass-estimate}.  Finally, we summarize the results
obtained in Sect.~\ref{sec:conclusions}.

A few plots in this paper are associated with switches designed to
show or hide labels; these are shown as frames in the respective
captions.  In order to use this feature this electronic document
should be displayed using Adobe$\textsuperscript{\textregistered}$
Acrobat$\textsuperscript{\textregistered}$.

\section{N{\small ICER} and N{\small ICEST} extinction maps}
\label{sec:nicer-absorpt-map}

\begin{figure*}[!tbp]
  \centering
  \includegraphics[angle=90, width=15cm]{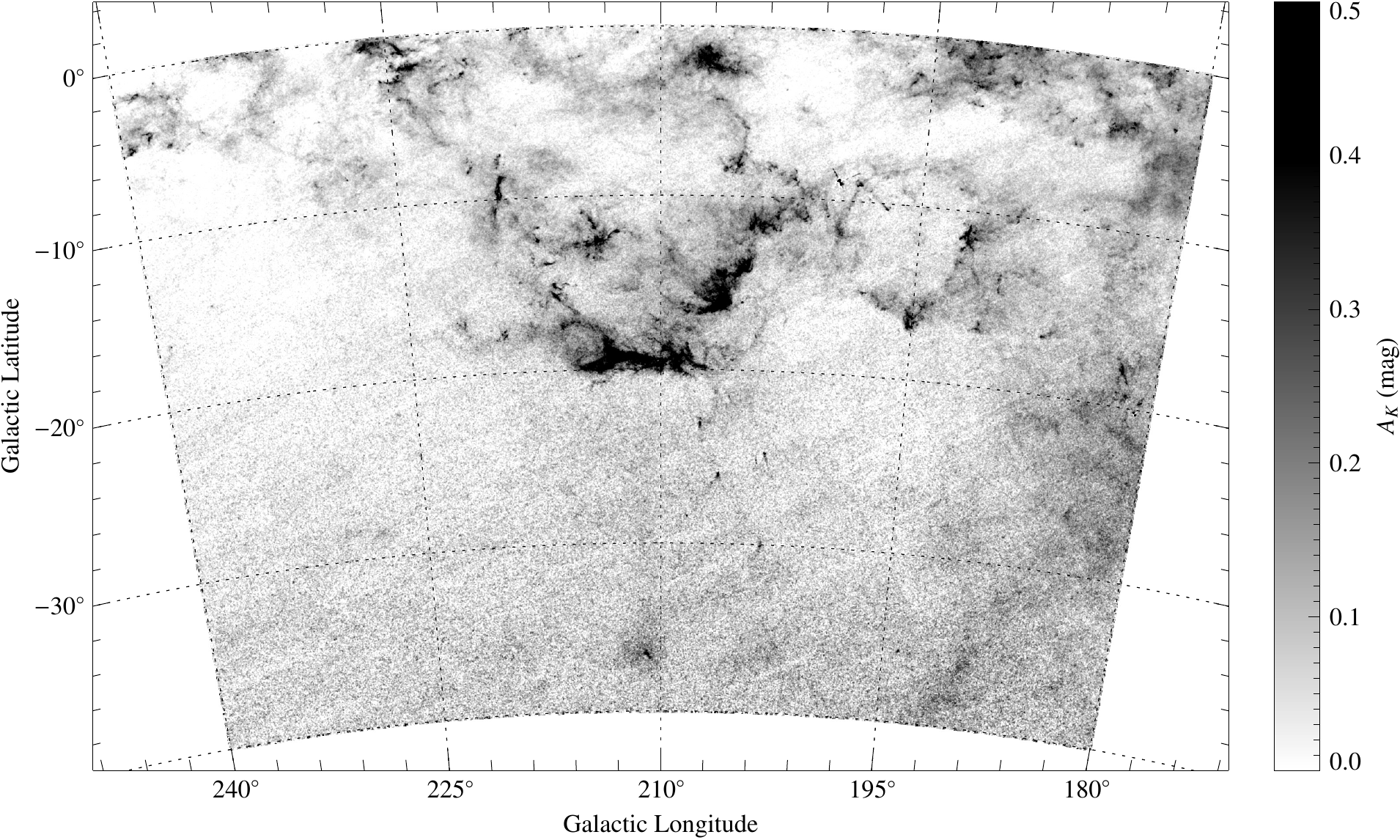}%
  \hspace{-15cm}%
  \begin{ocg}{fig3}{3}{1}%
    \includegraphics[angle=90, width=15cm]{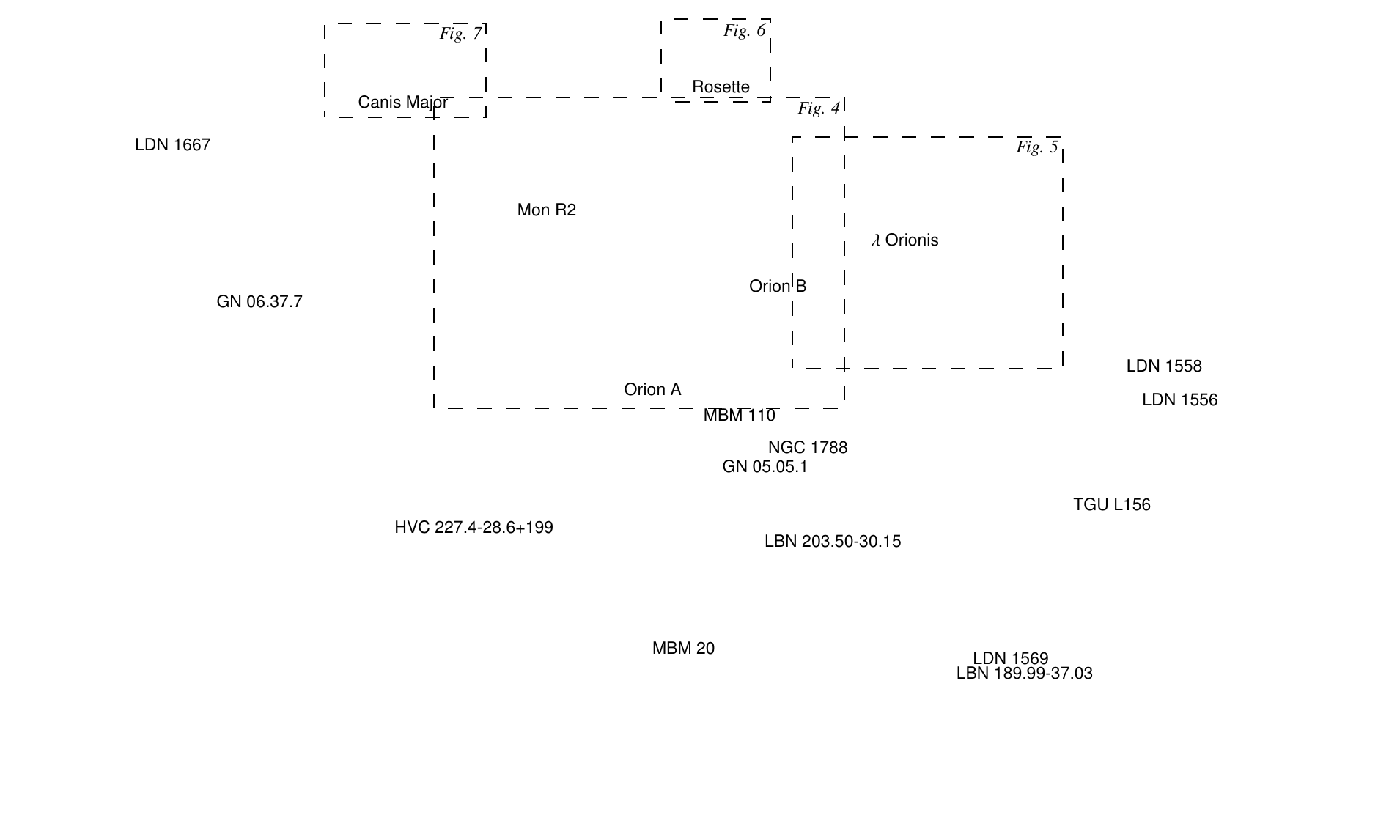}%
  \end{ocg}%
  \caption{The \textsc{Nicer} extinction map of the Orion, Mon R2,
    Rosette, and Canis Major complexes.  The resolution is
    $\mathrm{FWHM} = 3 \mbox{ arcmin}$.  The various dashed boxes
    mark the regions shown in greater detail in
    Figs.~\ref{fig:4}--\ref{fig:7}.  Toggle
    \ToggleLayer{fig3}{\protect\framebox{labels}}.}
  \label{fig:3}
\end{figure*}

The data analysis was carried out following the technique presented in
Paper~0 and used also in the previous papers of this series, to which
we refer for the details (see in particular Paper~III).  We selected
reliable point source detections from the Two Micron All Sky
Survey\footnote{See \texttt{http://www.ipac.caltech.edu/2mass/}.}
\citep[2MASS;][]{1994ExA.....3...65K} in the region
\begin{align}
  \label{eq:1}
  180^\circ <{} & l < 240^\circ \; , & -40^\circ <{} & b < 0^\circ \; .
\end{align}
This area is $\sim 2\,200$ square degrees and contains approximately
$19$ million point sources from the 2MASS catalog.  The region
encloses many known dark molecular cloud complexes, including the
Orion and Mon R2 star forming regions, the $\lambda$ Orionis bubble,
the Rosette nebula, and the Canis Major complex (see Fig.~\ref{fig:3}).

\begin{figure}[!t]
  \centering
  \includegraphics[width=\hsize]{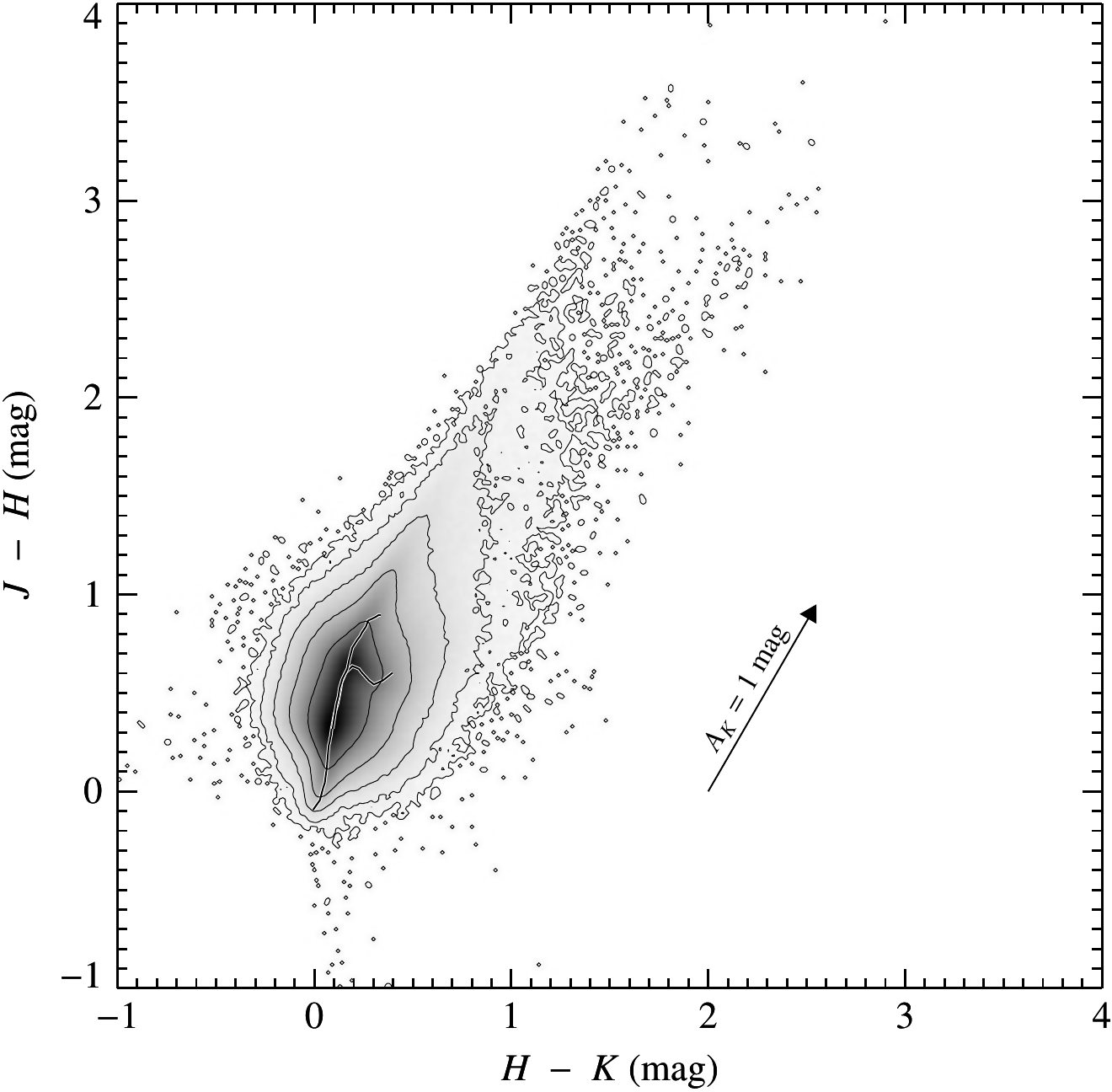}
  \caption{Color-color diagram of the stars in the whole field.  The
    contours are logarithmically spaced, with each contour
    representing a density ten times larger than the enclosing
    contour; the outer contour detects single stars.  Only stars with
    accurate photometry in all bands (maximum 1-$\sigma$ errors
    allowed $0.1 \mbox{ mag}$) have been included in this plot.  The
    diagonal spread of stars is due to reddening along the marked
    reddening vector of dwarf and giant stars.  The expected colors of
    these stars, as predicted by \citet{1988PASP..100.1134B} and
    converted into the 2MASS photometric system using the relations
    from \citep{2001AJ....121.2851C}, are shown in the plot, and match
    very well the observed colors.}
  \label{fig:2}
\end{figure}

\begin{figure*}[!tbp]
  \centering
  \includegraphics[width=\hsize]{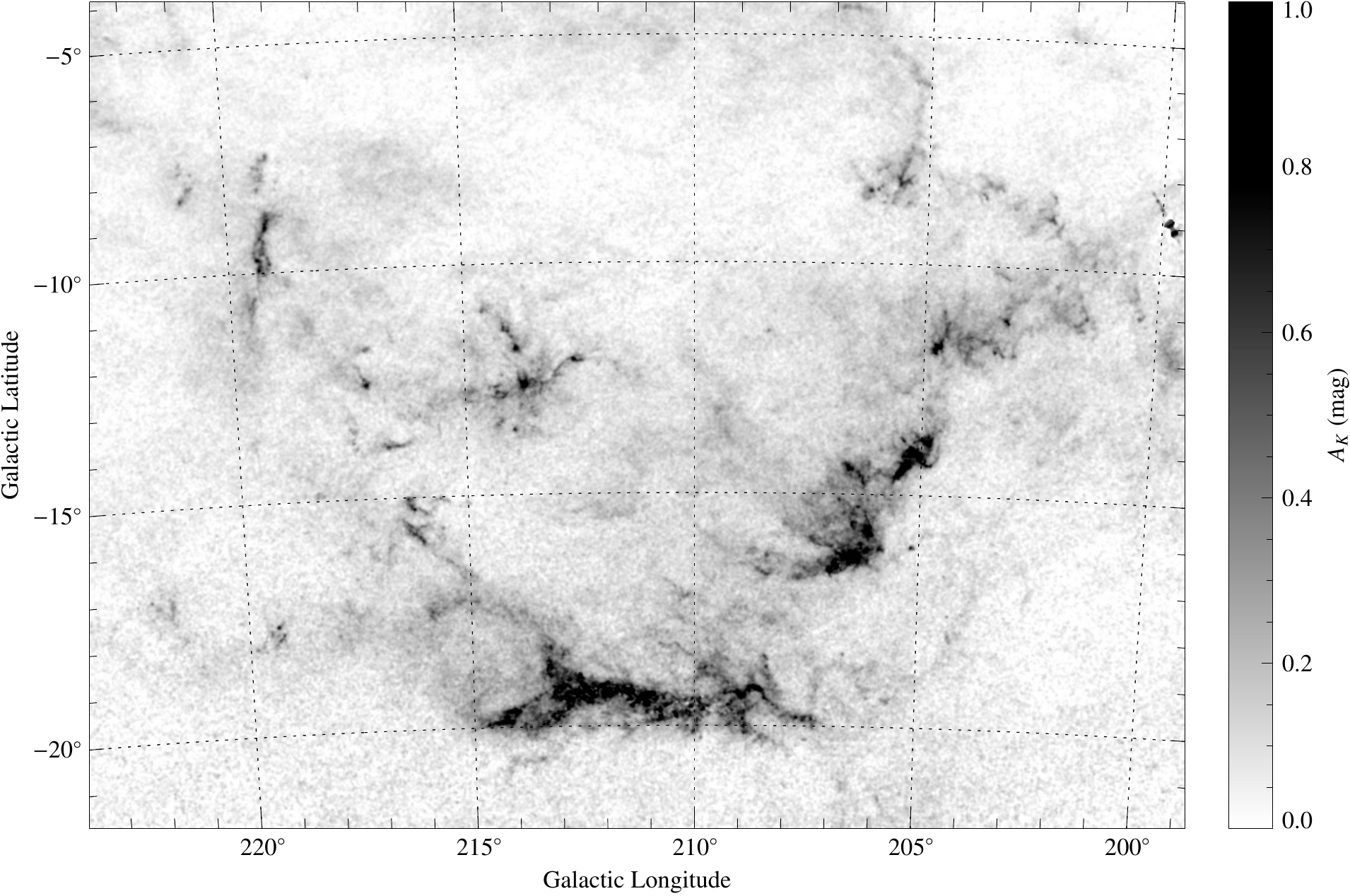}%
  \hspace{-\hsize}%
  \begin{ocg}{fig4a}{4}{1}%
    \includegraphics[width=\hsize]{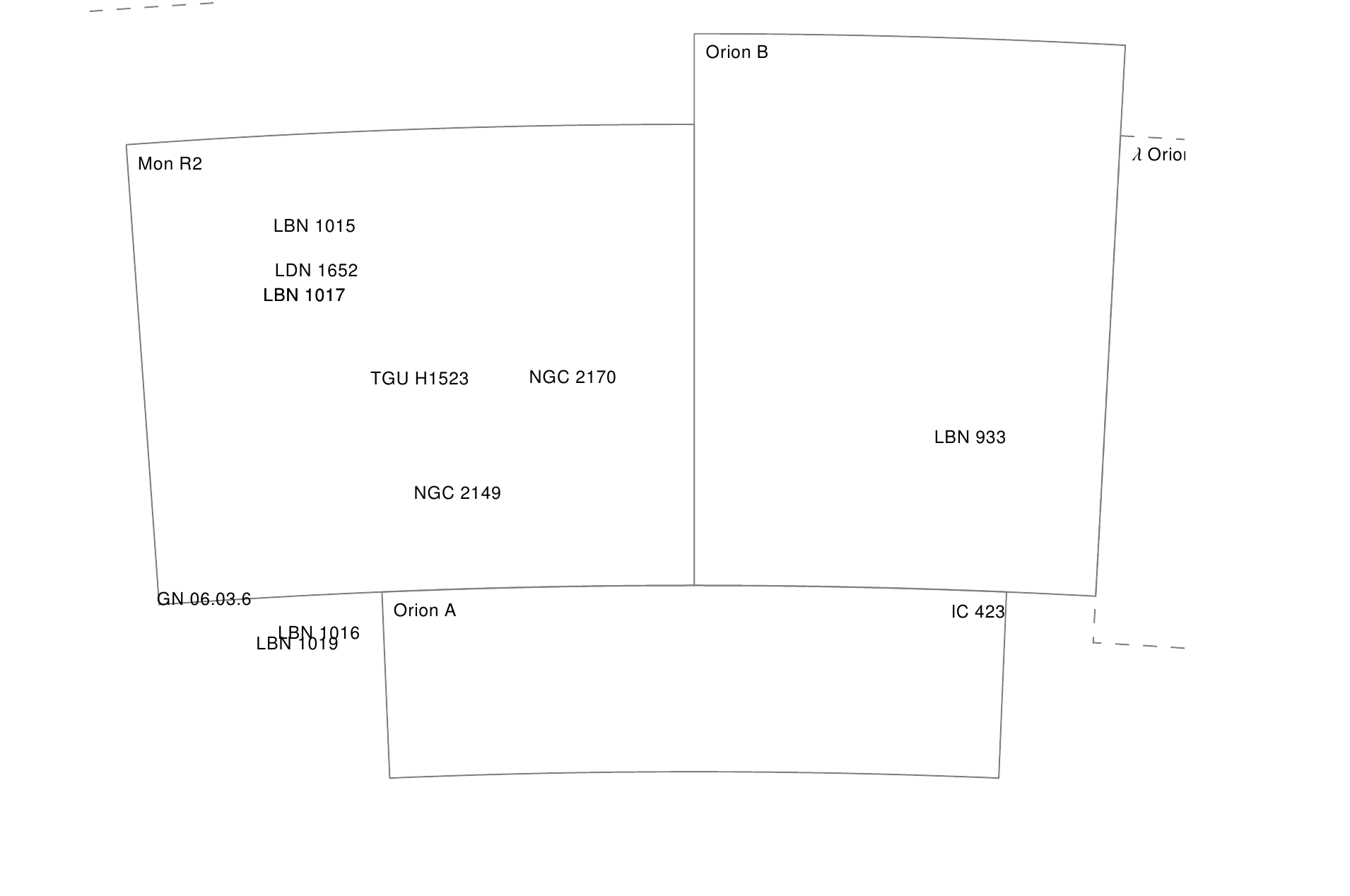}%
  \end{ocg}%
  \caption{A zoom of Fig.~\ref{fig:3} showing the Orion A (bottom),
    Orion B (top right), and Mon R2 (top left) star forming regions.
    The several well studied objects are marked.  The artifacts around
    $l = 200^\circ$ and $b = -9^\circ$ are caused by Betelgeuse.
    Toggle \ToggleLayer{fig4a}{\protect\framebox{labels}}.}
  \label{fig:4}
\end{figure*}

\begin{figure}[!tbp]
  \centering
  \includegraphics[width=\hsize]{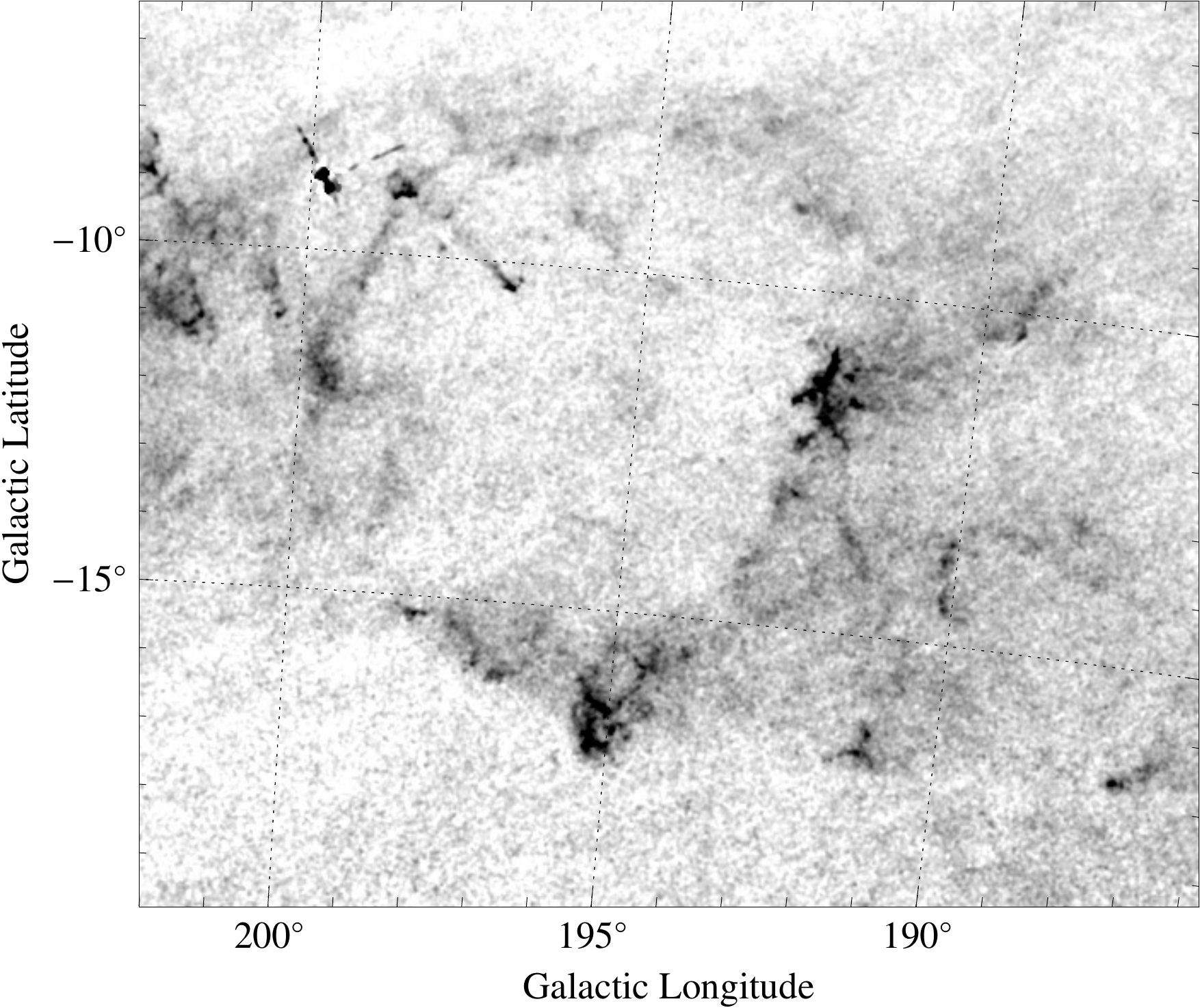}%
  \hspace{-\hsize}%
  \begin{ocg}{fig4b}{5}{1}%
    \includegraphics[width=\hsize]{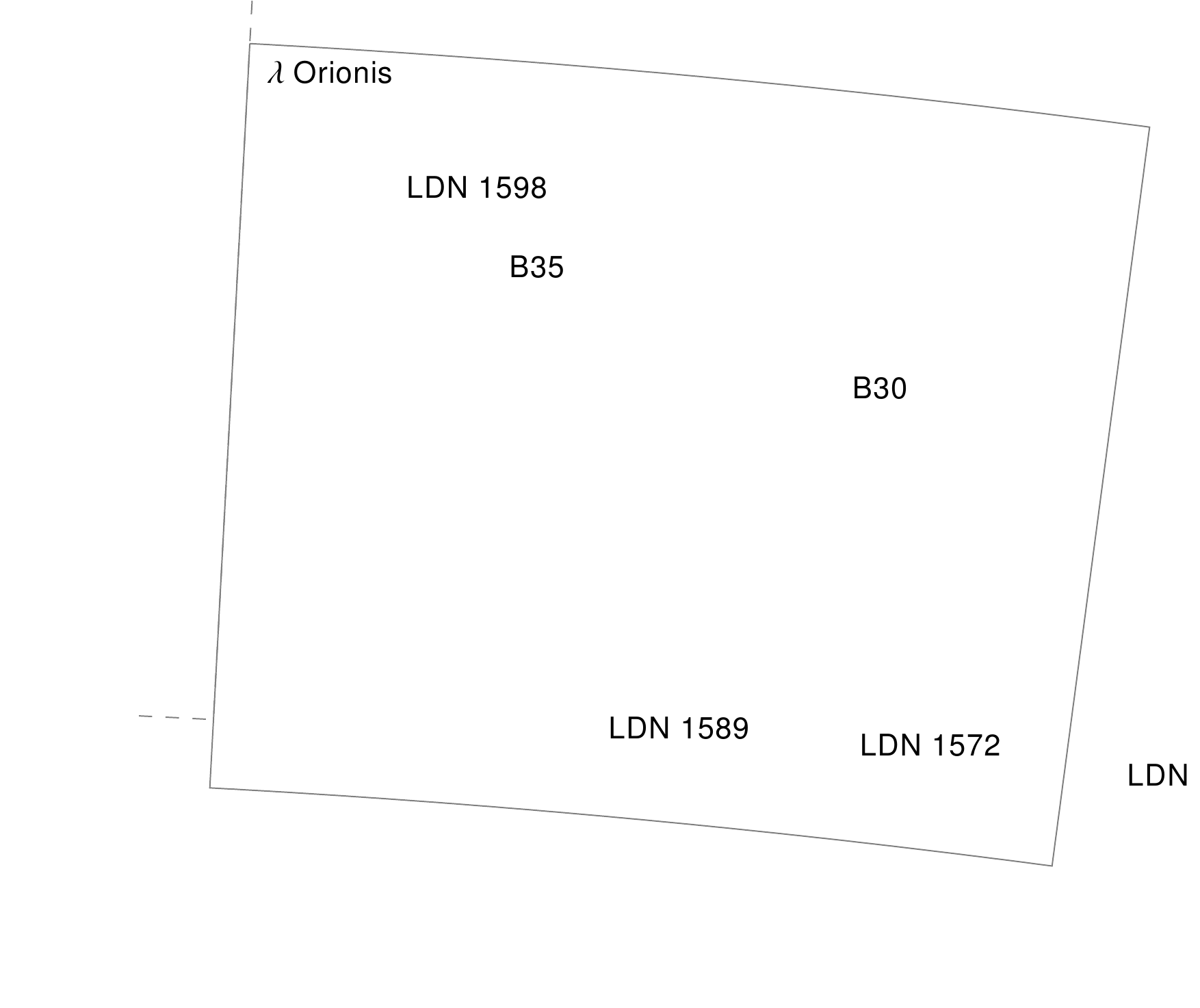}
  \end{ocg}%
  \caption{A zoom of Fig.~\ref{fig:3} showing the $\lambda$ Orionis
    bubble.  Toggle \ToggleLayer{fig4b}{\protect\framebox{labels}}.}
  \label{fig:5}
\end{figure}

\begin{figure}[!tbp]
  \centering
  \includegraphics[width=\hsize]{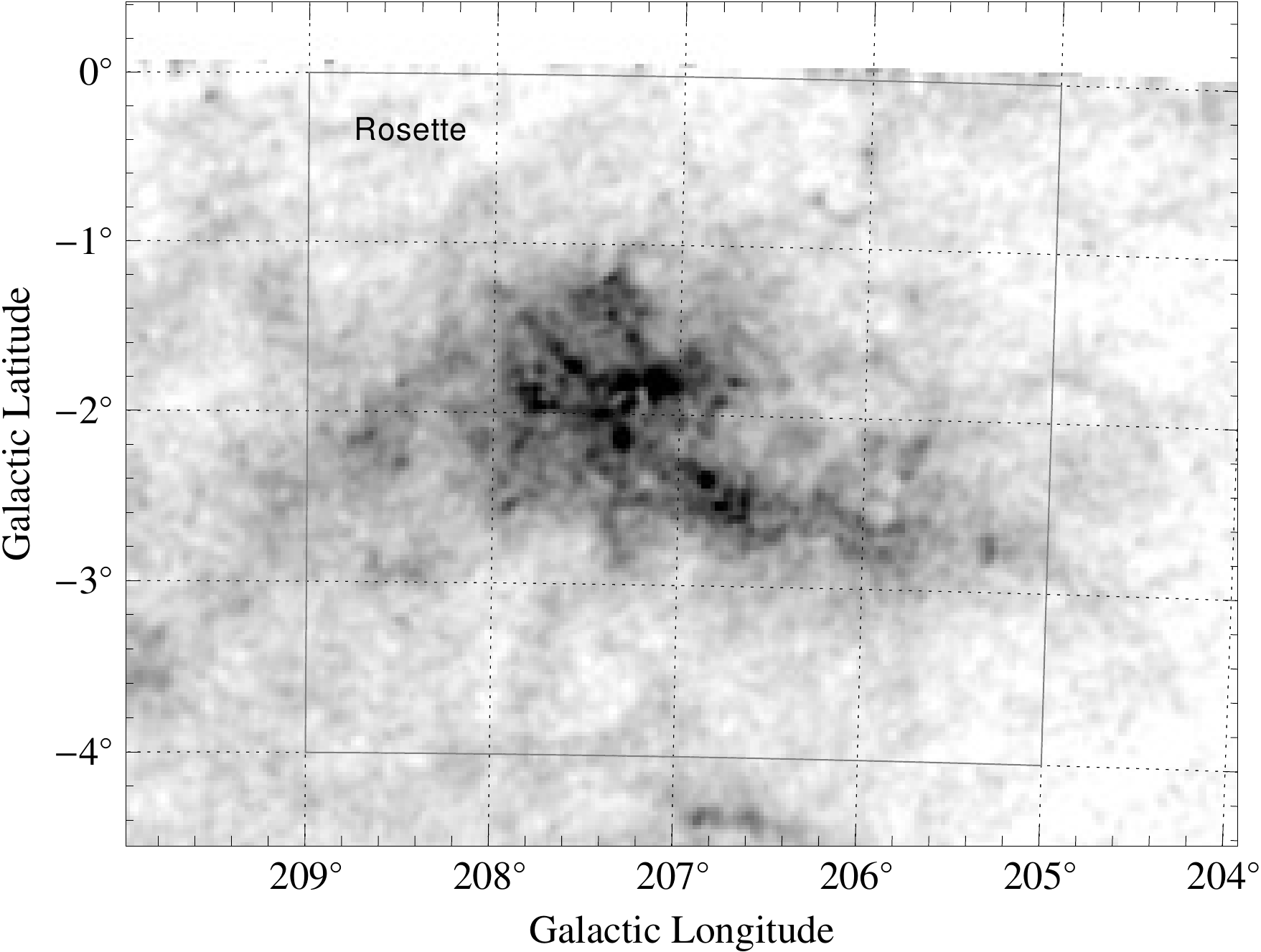}
  \caption{A zoom of Fig.~\ref{fig:3} showing the Rosette complex.}
  \label{fig:6}
\end{figure}

\begin{figure}[!tbp]
  \centering
  \includegraphics[width=\hsize]{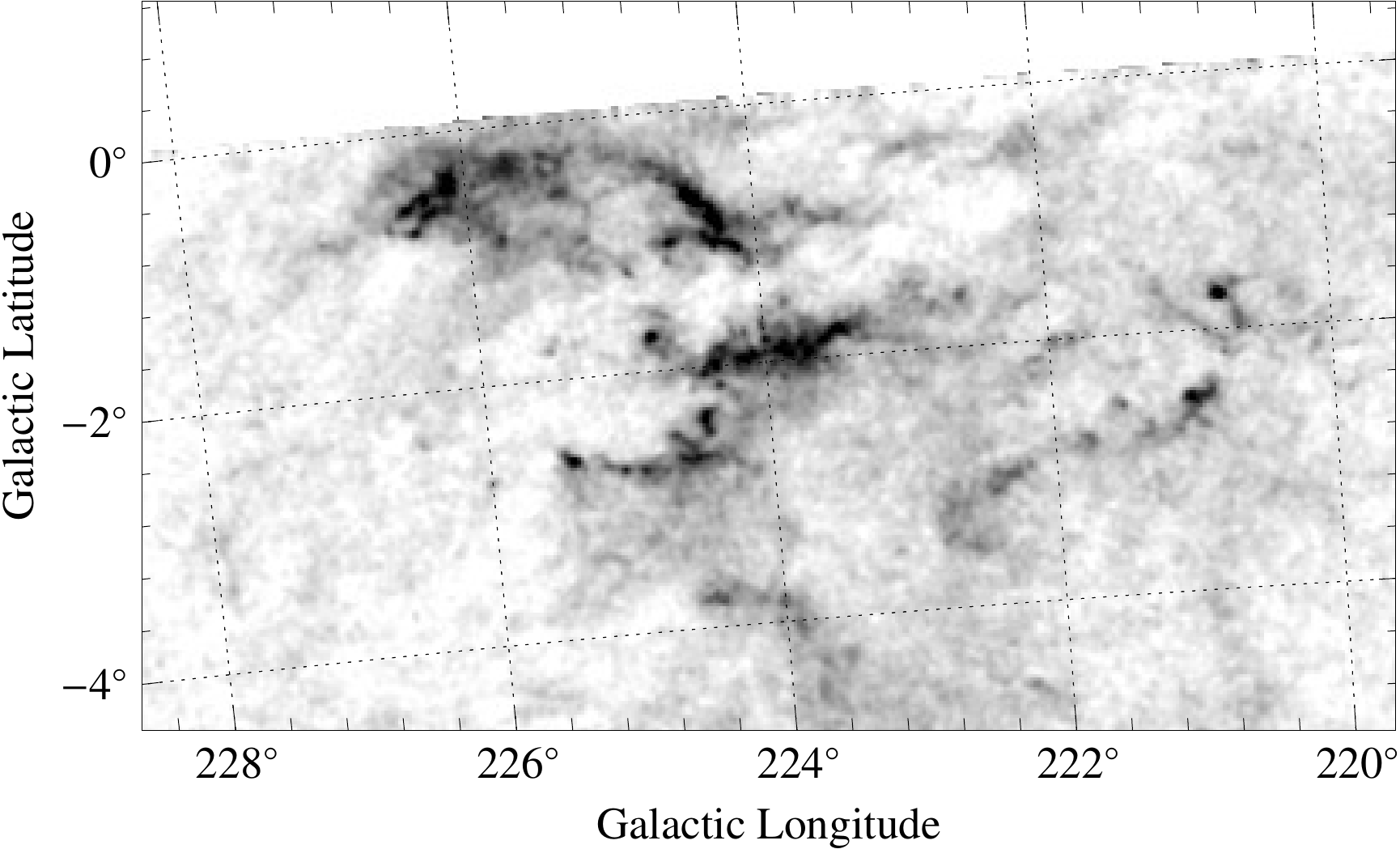}%
  \hspace{-\hsize}%
  \begin{ocg}{fig4d}{6}{1}%
    \includegraphics[width=\hsize]{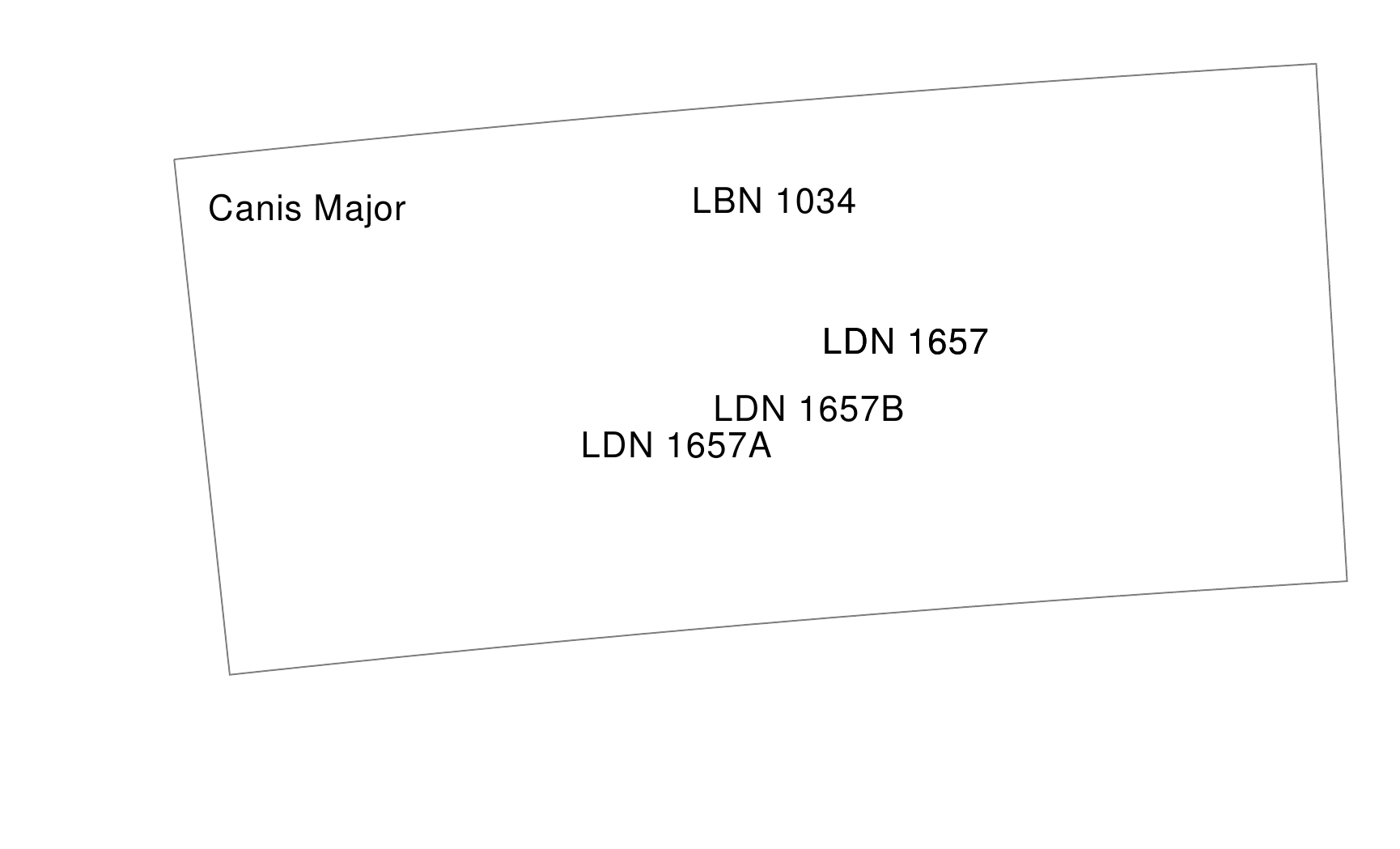}
  \end{ocg}%
  \caption{A zoom of Fig.~\ref{fig:3} showing the Canis Major dark
    complex.  Toggle \ToggleLayer{fig4d}{\protect\framebox{labels}}.}
  \label{fig:7}
\end{figure}

As a preliminary step, we constructed the color-color diagram of the
stars to check for the possible presence of anomalies in colors of
stars.  The result, shown in Fig.~\ref{fig:2}, displays a weak sign of
bifurcation in the distribution of source colors along the reddening
vector.  This is likely to be due to the different colors of
extinguished field stars and Asymptotic Giant Branch (AGB) stars.
However, given the weakness of the AGB stars contamination, we decided
to proceed similarly to Paper~III and not exclude any object.

After the selection of a control field for the calibration of the
intrinsic colors of stars (and their covariance matrix) we produced
the final 2MASS/\textsc{Nicer} extinction map, shown in
Fig.~\ref{fig:3}.  To obtain these results, we smoothed the individual
extinctions measured for each star, $\bigl\{ \hat A^{(n)}_K \bigr\}$,
using a moving weighted average
\begin{equation}
  \label{eq:2}
  \hat A_K(\vec\theta) = \frac{\sum_{n=1}^N W^{(n)}(\vec\theta) \hat
    A^{(n)}_K }{\sum_{n=1}^N W^{(n)}(\vec\theta)} \; ,
\end{equation}
where $\hat A_K(\vec\theta)$ is the extinction at the angular position
$\vec\theta$ and $W^{(n)}(\vec\theta)$ is the weight for the $n$-th
star for the pixel at the location $\vec\theta$.  This weight, in the
standard \textsc{Nicer} algorithm, is a combination of a smoothing,
window function $W\bigl( \vec\theta - \vec\theta^{(n)} \bigr)$, i.e.\
a function of the angular distance between the star and the point
$\vec\theta$ where the extinction has to be interpolated, and the
inverse of the inferred variance on the estimate of $A_K$ from the
star:
\begin{equation}
  \label{eq:3}
  W^{(n)}(\vec\theta) = \frac{W \bigl( \vec\theta - \vec\theta^{(n)}
    \bigr)}{\mathrm{Var}\bigl( \hat A^{(n)}_K \bigr)} \; .
\end{equation}
Note that the way Eq.~\eqref{eq:2} is written, only
\textit{relative\/} values of $W^{(n)}$, and thus of $W\bigl(
\vec\theta - \vec\theta^{(n)} \bigr)$ are important.  Therefore, we
can assume without loss of generality that the window function is
normalized to unity according to the equation
\begin{equation}
  \label{eq:4}
  \int W(\vec\theta) \, \diff^2 \theta = 1 \; .
\end{equation}
For this paper, the smoothing window function $W$ was taken to be a
Gaussian with $\mathit{FWHM} = 3 \mbox{ arcmin}$.  Finally, the map
described by Eq.~\eqref{eq:2} was sampled at $1.5 \mbox{ arcmin}$
(corresponding to a Nyquist frequency for the chosen window function).

Similarly to Paper~III, we also constructed a \textsc{Nicest}
extinction map, obtained by using the modified estimator described in
\citet{2009A&A...493..735L}.  The \textsc{Nicest} map differs
significantly from the \textsc{Nicer} map only in the high
column-density regions, where the substructures present in the
molecular cloud produce a possibly significant bias on the standard
estimate of the column density (see below).  The largest extinction
was measured close to LDN~1641~S, where $A_K \simeq 2.0 \mbox{
  mag}$ for \textsc{Nicer} and $A_K \simeq 5.0 \mbox{ mag}$ for
\textsc{Nicest}.

Figures~\ref{fig:4}--\ref{fig:7} show in greater detail the absorption
maps we obtained for the Orion and Mon R2 star forming regions, the
$\lambda$ Orionis bubble, the Rosette nebula, and the Canis Major
complex.  These maps allow us to better appreciate the details that we
can obtain by applying the \textsc{Nicer} method to the 2MASS data.
In the figures we also display the boundaries that we use throughout
this paper and that we associate with the various clouds considered
here.  In particular, we defined
\begin{align}
  \label{eq:5}
  & \text{Orion A:} &
  203^\circ \le l & {} \le 217^\circ \; , &
  -21^\circ \le b & {} \le -17^\circ \; , \notag\\
  & \text{Orion B:} &
  201^\circ \le l & {} \le 210^\circ \; , &
  -17^\circ \le b & {} \le -5^\circ \; , \notag\\
  & \text{Mon R2:} &
  210^\circ \le l & {} \le 222^\circ \; , &
  -17^\circ \le b & {} \le -7^\circ \; , \notag\\
  & \text{$\lambda$ Orionis:} &
  188^\circ \le l & {} \le 201^\circ \; , &
  -18^\circ \le b & {} \le -7^\circ \; , \notag\\
  & \text{Rosette:} &
  205^\circ \le l & {} \le 209^\circ \; , &
  -4^\circ \le b & {} \le 0^\circ \; , \notag\\
  & \text{Canis Major:} &
  220^\circ \le l & {} \le 228^\circ \; , &
  -4^\circ \le b & {} \le 0^\circ \; .
\end{align}
We note that the boundaries defined are somewhat arbitrary, but as
described below (see Sect.~\ref{sec:foregr-star-cont} and in
particular Fig.~\ref{fig:9}) there are strong indications that many
(if not all) of the features within each of the defined regions are
located at similar distances.  We stress in any case that some of the
results presented (for example, the mass estimates discussed in
Sect.~\ref{sec:mass-estimate}) depend significantly on the chosen
boundaries.

The expected error on the measured extinction, $\sigma_{\hat A_K}$
(not shown here), was evaluated from a standard error propagation in
Eq.~\eqref{eq:2} [see below Eq.~\eqref{eq:9}].  The main factor
affecting the error is the local density of stars, and thus there is a
significant change in the statistical error along galactic latitude.
Other variations can be associated with bright stars (which are masked
out in the 2MASS release), bright galaxies, and to the cloud itself.
The median error per pixel for the field shown in Fig.~\ref{fig:4} is
$0.029 \mbox{ mag}$, while a significantly lower error of $0.019
\mbox{ mag}$ is observed for the Rosette or Canis Major nebul\ae.

\section{Statistical analysis}
\label{sec:statistical-analysis}

\subsection{Reddening law}
\label{sec:reddening-law}

\begin{figure*}[!tbp]
  \begin{center}
    \includegraphics[width=\hsize]{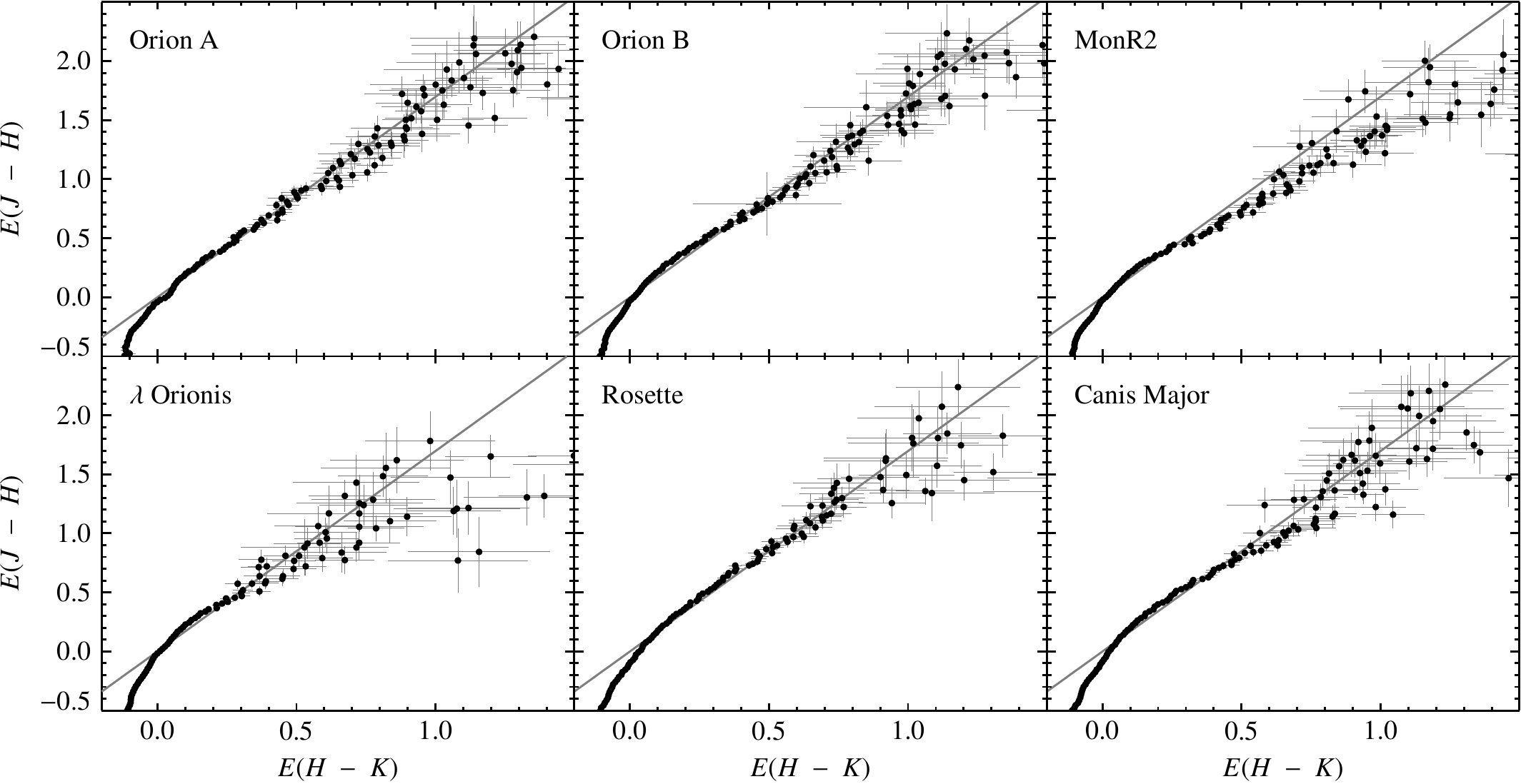}
    \caption{The reddening law as measured on various complexes.  The
      plots show the color excess on $J - H$ as a function of the
      color excess on $H - K$.  Error bars are uncertainties evaluated
      from the photometric errors of the 2MASS catalog and taking into
      account also the intrinsic scatter in the individual colors of
      the stars, as derived from the control field.  The solid line
      shows the normal infrared reddening law
      \citep{2005ApJ...619..931I}.  A significant discrepancy is
      observed for Mon R2 and possibly for $\lambda$ Orionis, while
      other clouds are in very good agreement with the normal infrared
      reddening law.  }
    \label{fig:8}
  \end{center}
\end{figure*}

As shown in Paper~0, the use of multiple colors in the \textsc{Nicer}
algorithm significantly improves the signal-to-noise ratio of the
final extinction maps, but also provides a simple, direct way to
verify the reddening law.  A good way to perform this check is to
divide all stars with reliable measurements in all bands into
different bins corresponding to the individual $\hat A_K^{(n)}$
measurements, and to evaluate the average NIR colors of the stars in
the same bin.  At first, this procedure might be regarded as a
circular argument: in order to measure the extinction law one bins the
colors of stars according to the estimated extinction!  In reality,
the method is well behaved and has interesting properties (Ascenso et
al. 2011, in preparation):
\begin{itemize}
\item The assumed reddening law is only used to bin the data, and is
  then iteratively replaced by the new reddening law as determined
  from the data.
\item This iterative process converges quickly and is essentially
  unbiased; the final reddening law essentially does not depend on the
  initial assumed one.
\item Simulations show that more standard techniques (such as binning
  in one simple color, e.g.\ $J - K$) suffer from significant biases
  mostly as a result of the heteroskedasticity of the data; additional
  complications are due to the correlation of errors in the two colors
  $J-H$ and $H-K$.
\end{itemize}

Figures~\ref{fig:8} summarize the results obtained for a bin size of
$0.02 \mbox{ mag}$ in the various clouds.  As shown by these plots, in
all clouds we have a good agreement between the standard
\citet{2005ApJ...619..931I} infrared reddening law in the 2MASS
photometric system and the observed one.  All plots show a systematic
divergence below the reddening line at ``negative'' extinctions, an
effect due to the intrinsic colors of dwarf stars
\citep[see][]{1983A&A...128...84K,1988PASP..100.1134B}.  Specifically,
the track of dwarf stars in the $J - H$ vs.\ $H-K$ color-color plot is
slightly steeper than the reddening vector (cf.\ Fig.~\ref{fig:2}),
and as a result stars in the bottom-left of the dwarf track, which in
the simple \textsc{Nicer} scheme are associated to negative
extinction, exhibit an excess in their $H-K$ color.  This effect is
not visible anymore as we go to positive extinction because the colors
of these stars are then averaged together with the rest of the dwarf
and giant sequence, and as imposed by the control field no systematic
color excesses is present.

Another common feature to all plots of Fig.~\ref{fig:8} is an increase
in the spread of points in the upper-right corner, i.e.\ for high
column densities.  As shown by the correspondingly larger error bars,
this is expected and is a result of the low number of stars showing
large extinction values (we recall that a constant bin size of $0.02
\mbox{ mag}$ has been used everywhere).  Additionally, for some
clouds, and in particular for Mon~R2, we observe that the points fall
systematically below the standard \citet{2005ApJ...619..931I}
reddening law.  We believe that this difference is \textit{not\/} due
to intrinsic differences in the physical properties of the dust, but
rather that it is caused by contamination from a population of
relatively blue, young stars present in the field, both in the form of
clusters and of dispersed OB associations (in particular, the area is
known to host the Mon OB1 association, see
\citealp{1968AJ.....73..233R, 1976AJ.....81..840H,
  2008hsf1.book..899C}).

Finally, we stress that the check performed with Fig.~\ref{fig:8} is a
\textit{relative\/} one: we can only verify that slope of the
reddening vector agrees with expectations from the standard
\citet{2005ApJ...619..931I} reddening law, but can not constrain the
length of the reddening vector.

\begin{figure*}[!tbp]
  \begin{center}
    \includegraphics[width=\hsize]{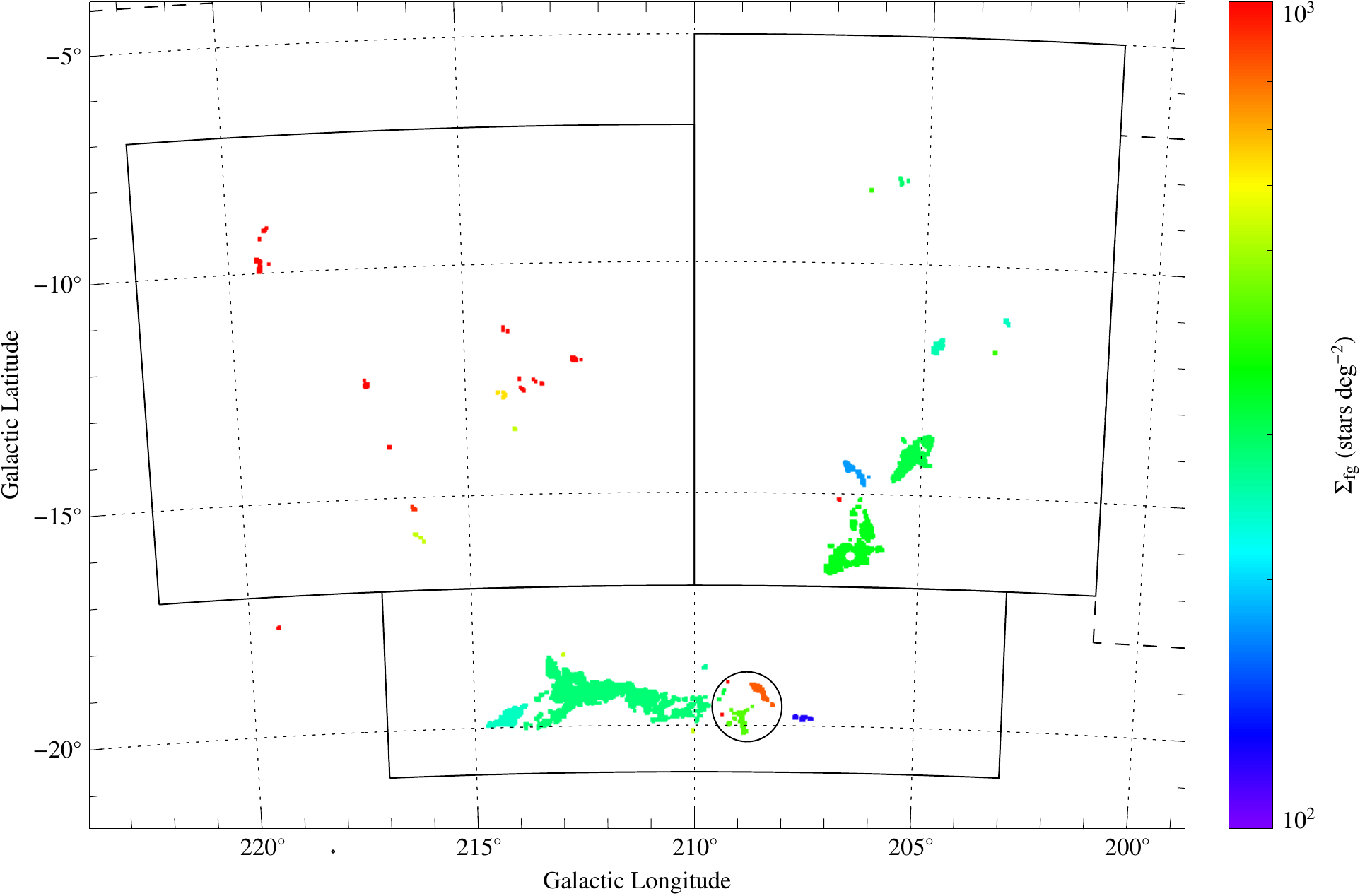}
    \caption{The local density of foreground stars, averaged on
      connected regions with extinction $A_K > 0.6 \mbox{ mag}$.  The
      three regions defined in Fig.~\ref{fig:4} are shown again here.}
    \label{fig:9}
  \end{center}
\end{figure*}

\begin{figure*}[!tbp]
  \begin{center}
    \includegraphics[width=0.32\hsize]{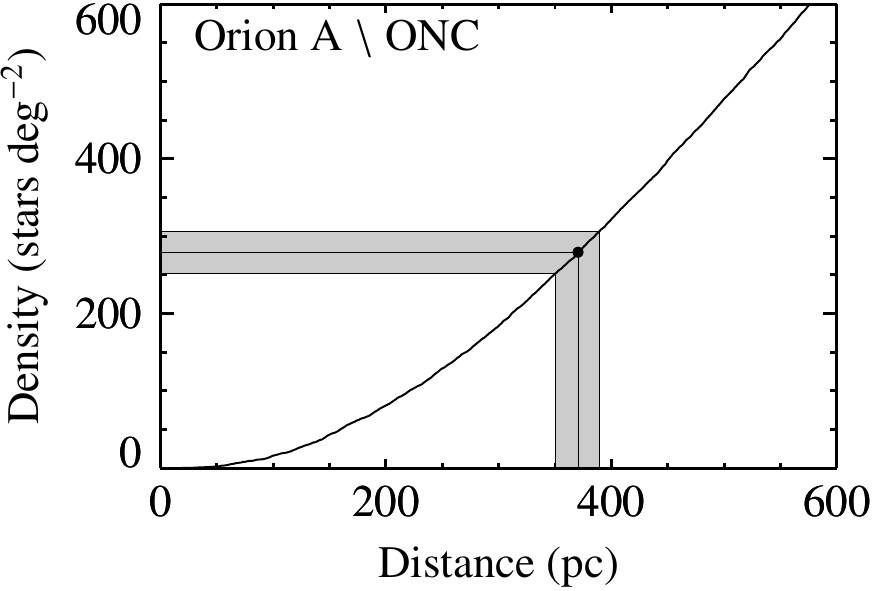}
    \includegraphics[width=0.32\hsize]{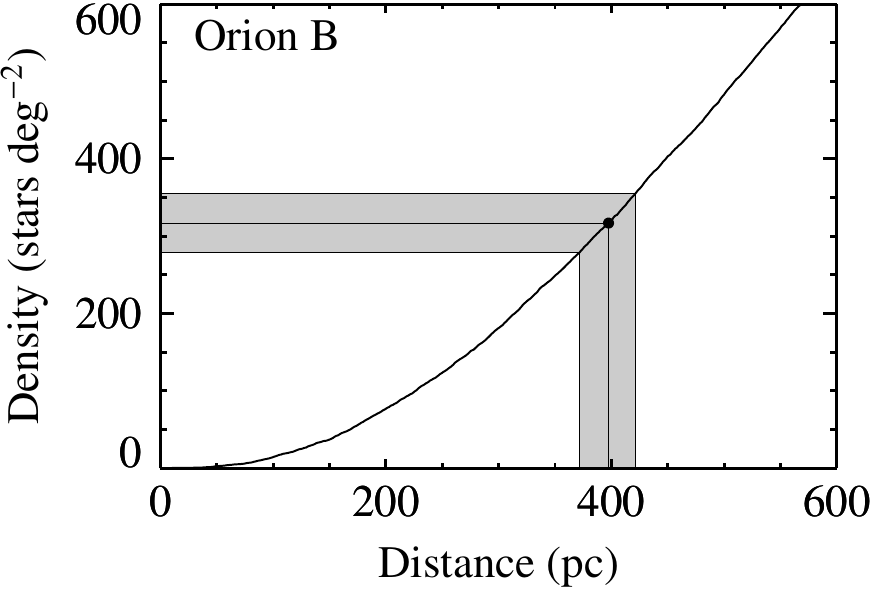}
    \includegraphics[width=0.32\hsize]{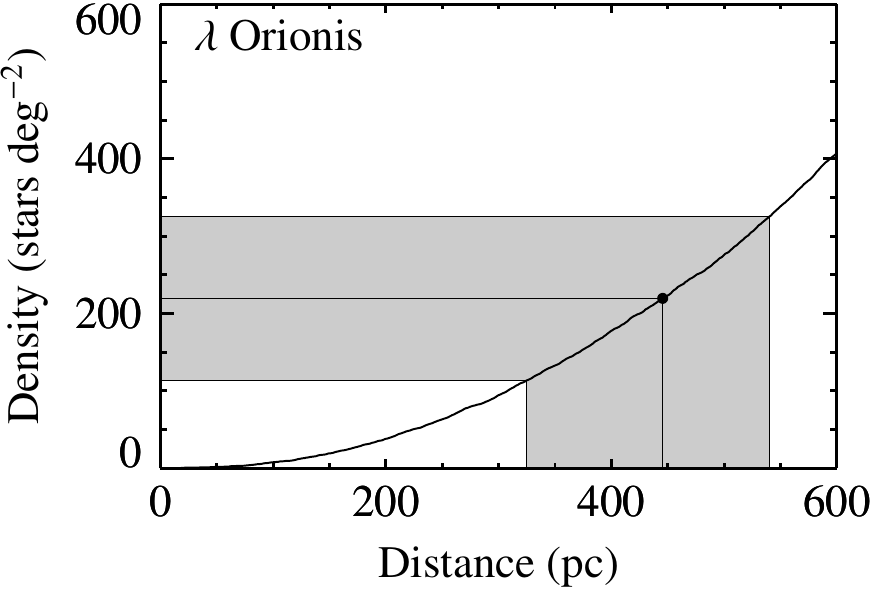}
    \includegraphics[width=0.32\hsize]{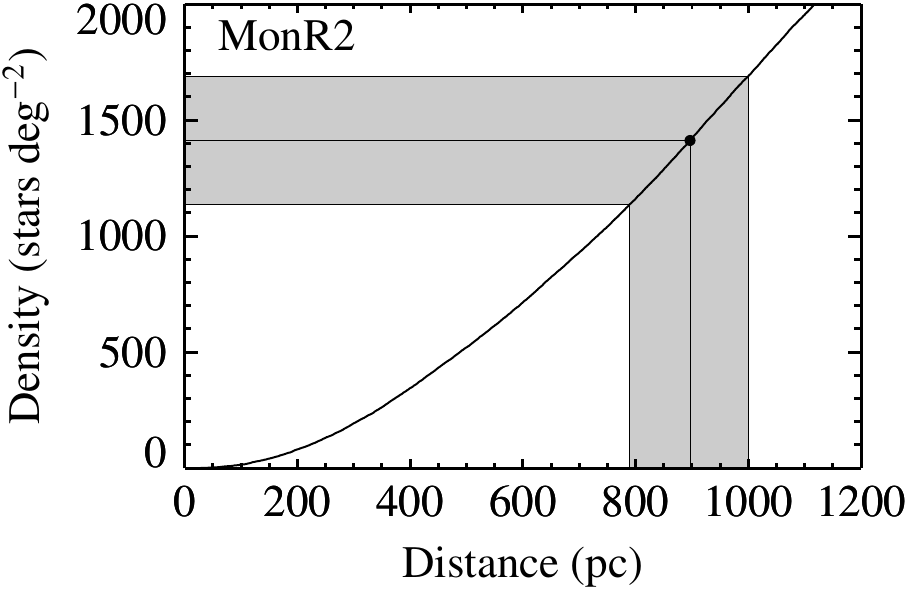}
    \includegraphics[width=0.32\hsize]{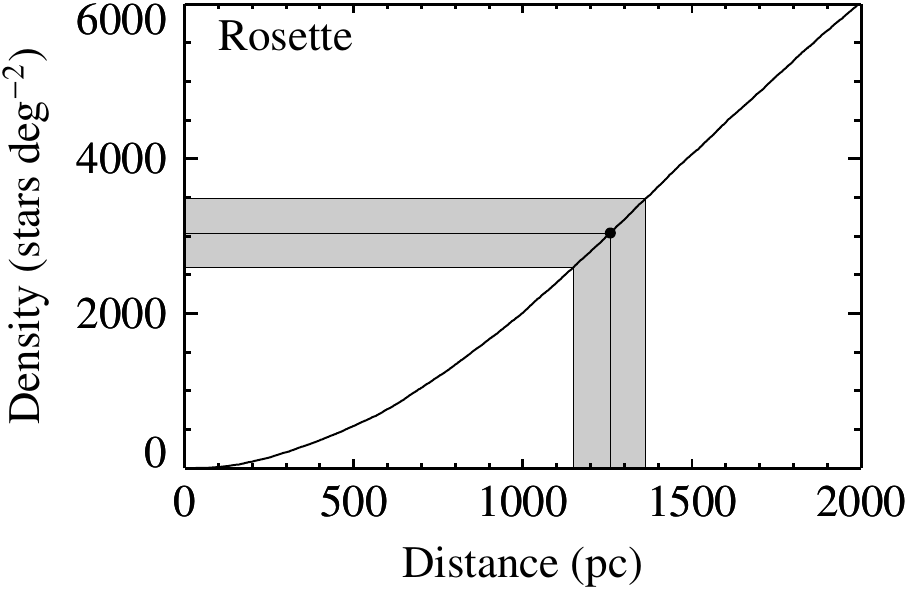}
    \includegraphics[width=0.32\hsize]{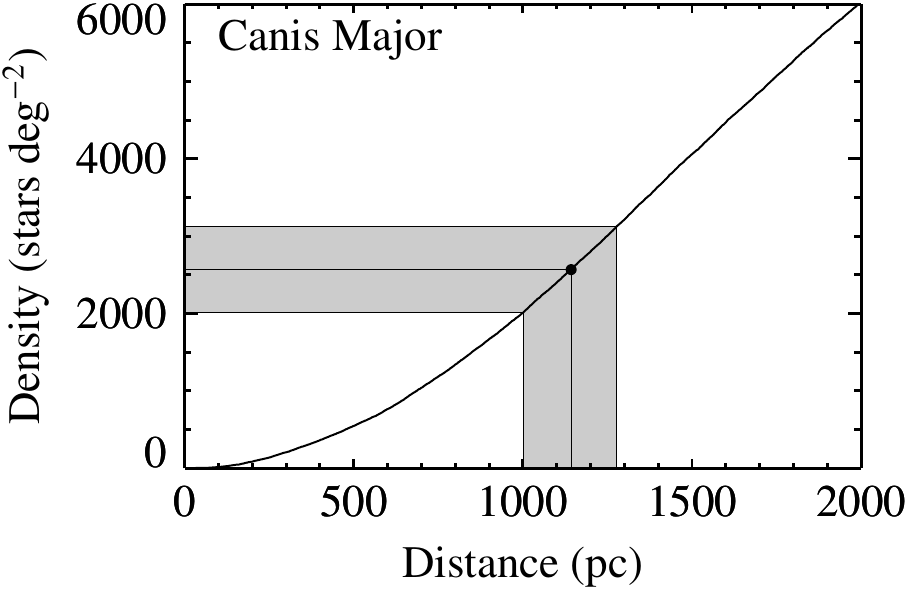}
    \caption{The distances of the clouds deduced from the density of
      foreground stars.  Solid lines show the density of foreground of
      stars as a function of the cloud distance, as predicted from the
      \citet{2003A&A...409..523R} Galactic model.  The grey areas
      correspond to the $95\%$ (two-sigma) confidence regions for the
      density of foreground stars and the deduced confidence regions
      for the distances.}
    \label{fig:10}
  \end{center}
\end{figure*}

\subsection{Foreground star contamination and distances}
\label{sec:foregr-star-cont}

Foreground stars are an annoyance for color extinction studies, since
they dilute the signal coming from background stars and add a source
of noise to the maps.  For nearby clouds, foreground stars are usually
a small fraction of the total number of stars in all regions except in
the very dense cores (where the density of background stars decreases
significantly because of the dust extinction).  In these conditions,
one can safely ignore the bias introduced by foreground stars, and if
necessary use \textsc{Nicest} measurements to alleviate the foreground
bias in the high column-density regions.

\begin{table}[b!]
  \def\0{\phantom{0}}
  \centering
  \begin{tabular}{lccccc}
    Complex & $N_\mathrm{fg}$ & $f$ & Area &
    $\Sigma_\mathrm{fg}$ & Distance \\
    & & & $\mbox{deg}^2$ & $\mbox{deg}^{-2}$ & pc \\
    \hline
    Orion A       & 420 & 0.047 & 1.504 & $\0279 \pm 13\0$ & $\0371 \pm 10$ \\
    Orion B       & 298 & 0.030 & 0.936 & $\0318 \pm 18\0$ & $\0398 \pm 12$ \\
    $\lambda$ Orionis 
        & \phantom{0}17 & 0.018 & 0.078 & $\0219 \pm 53\0$ & $\0445 \pm 50$ \\
    Mon R2        & 207 & 0.157 & 0.144 & $ 1430 \pm 99\0$ & $\0905 \pm 37$ \\
    Rosette       & 275 & 0.190 & 0.083 & $ 3330 \pm 200 $ & $ 1330 \pm 48$ \\
    Canis Major   & 
          \phantom{0}99 & 0.228 & 0.036 & $ 2730 \pm 270 $ & $ 1150 \pm 64$
  \end{tabular}
  \caption{The average value of foreground stars $N_\mathrm{fg}$ and
    their fraction $f$ found in the various
    complexes.  Also reported in the last column the estimated
    distance from a comparison with the \citet{2003A&A...409..523R}
    Galactic model (see Fig.~\ref{fig:10}).}
  \label{tab:1}
\end{table}

Foreground stars are easily selected in the dense regions of a
molecular cloud as objects showing no or very little sign of
extinction.  This allows us to estimate with good accuracy the local
density of foreground stars in all high-column density clouds in our
field.  For this purpose, we selected connected regions with $\hat
A_K(\vec\theta) > 0.6 \mbox{ mag}$ in our extinction map, and we
computed for each of these the local density of foreground stars
(foreground objects were selected as stars with K-band extinction
smaller than $0.3 \mbox{ mag}$, i.e.\ stars compatible with no or
negligible extinction).  Figure~\ref{fig:9} shows the results obtained
in Orion and Mon R2; the average numerical values obtained in the
various regions considered in this paper are also reported in
Table~\ref{tab:1}.  Two simple results are immediately visible from
Fig.~\ref{fig:9}.  First, the Orion cloud shows significantly fewer
foreground stars than the Mon R2 complex, as expected from the much
larger distance assigned in the literature to Mon R2.  Second, all
complexes have relatively uniform densities of foreground stars in
their regions except the area around the Orion Nebula Cluster (ONC),
marked with a circle in Fig.~\ref{fig:9}.  This result is also
expected, since it is well known that the ONC region contains a high
surface density of young stellar objects (YSOs) embedded in the cloud
that will contaminate our extinction measurements and a part of which
will be considered as foreground stars by our selection criteria.  In
order to obtain unbiased results, we excluded this region in the
analysis of the foreground star density.

We note that a comparison of the densities reported in
Table~\ref{tab:1} with the average density of stars in the field,
$\Sigma \sim 9\,000 \mbox{ deg}^{-2}$, shows that the expected
fraction $f$ of foreground stars in the outskirts of the Orion complex
is expected to be as low as $f \simeq 0.04$, while a significantly
larger value, $f \simeq 0.16$, is expected for Mon R2.

Interestingly, as shown in Paper~III, we can also use foreground stars
to estimate the distance of dark clouds.  The technique relies on a
comparison between the estimated density of foreground stars and the
predictions of galactic models for the photometric depth of the 2MASS
catalog.  We used the Galactic model by \citet{2003A&A...409..523R},
and computed at the location of each cloud the expected number of
stars within the 2MASS photometric limits observed at various
distances (see Fig.~\ref{fig:10}).  The results obtained, shown in the
last column of Table~\ref{tab:1}, presents several interesting
aspects.  First, the estimates for the Orion~A distance, when
excluding the highly contaminated area of the ONC, is $(371 \pm 10)
\mbox{ pc}$.  This value is considerably less than the ``standard''
Orion distance, $450 \mbox{ pc}$, but as discussed by
\citet{2008hsf1.book..483M}, this often quoted distance is actually
reported by \citet{1989ARA&A..27...41G} but it is not itself the
result of any direct measurement, but more likely an average of two
different measurements.  In general, in spite of the efforts over
several decades to measure the distance of this cloud complex, there
is still a spread in recent estimates at the $10\%$ level
\citep{2008hsf1.book..483M}.  We stress, however, that our measurement
is in good agreement with a recent VLBI determination, $(389 \pm 23)
\mbox{ pc}$ \citep{2007ApJ...667.1161S}, but lower than a measurement
by \citet{2007A&A...474..515M}, $(414 \pm 7) \mbox{ pc}$.

Our Orion~A distance is also very close to the Orion~B distance, a
result that is not unexpected given the physical relationship between
the two clouds.

We did attempt a measurement of the $\lambda$ Orionis distance, in
spite of the relative lack of dense material in the area.  This
paucity of high extinction material prevents us from estimating the
density of foreground stars with the necessary confidence.
Nonetheless, the result obtained is virtually identical to the
standard distance estimated by \citet{2001AJ....121.2124D} using
Stro\"omgren photometry of the OB stars in a main-sequence fitting in
a theoretical H-R diagram, $450 \pm 50 \mbox{ pc}$.  However, we are
unfortunately unable to improve this result in terms of accuracy.

The Mon R2 cloud is found to be at a significantly larger distance,
$(905 \pm 37) \mbox{ pc}$, a result that compares well with the
generally accepted distance of this cloud, $(830 \pm 50) \mbox{ pc}$
(\citealp{1976AJ.....81..840H}; see also
\citealp{2008hsf1.book..899C}).  We stress that, as evident from
Fig.~\ref{fig:9}, different clouds present in the MonR2 region seem to
have compatible surface densities of foreground stars and therefore
distances, with the possible exception of the filaments close to
NGC~2149.  In order to test quantitatively this statement, we also
considered subregions in the MonR2 cloud.  In particular, a
measurement of the distance of the LBN~1015, LDN~1652, and LBN~1017
clouds (the ``crossbones'') provides $(750 \pm 100) \mbox{ pc}$, while
a measurement of the MonR2 core alone (i.e.\ the area around NGC~2170)
gives $(925 \pm 150) \mbox{ pc}$.  These two regions therefore seem to
be at comparable distances, a result that is in contradiction with
what found by \citet{2005A&A...430..523W}.  Unfortunately, the data we
have are not sufficient to constrain the distance of the filaments
around NGC~2149, and therefore we cannot securely associate them with
MonR2 (nor with Orion~A).

The distance of the Rosette complex is generally obtained indirectly
by measuring the distance of the NGC~2244 cluster.  Typical values
obtained from photometric studies are around $1650 \mbox{ pc}$
\citep{1962ApJ...136.1135J, 1987PASP...99.1050P, 2002AJ....123..892P},
with the exception of \citet{1981PASJ...33..149O} that reports $1420
\mbox{ pc}$ (see also \citealp{2008hsf1.book..928R} for a discussion
of these measurements).  \citet{2000A&A...358..553H} performed a
spectroscopic analysis of the binary member V578~Mon, obtaining $1390
\mbox{ pc}$, a value that compares very well with our determination.

Recent distance estimates of the Canis Major region (and in particular
of its OB association) are close to $1000 \mbox{ pc}$ (see
\citealp{2008hsf2.book....1G} and references therein).  Although our
measurement appears to be slightly higher, the uncertainties of the
data present in the literature and of our own determination are
relatively large.

In summary, our analysis shows that for clouds outside the reach of
Hipparcos parallaxes, a model-dependent distance obtained through
number counts of foreground stars is a reasonable alternative of
distance determination.  Indeed, for all clouds studied here we
obtained results that are in very good agreement with the data present
in the literature (which often are obtained using dedicated
observations).  Additionally, the uncertainty we have is generally
comparable to the one that can be obtained with competitive techniques
(with the notable exception of VLBI parallaxes, which however are very
expensive in terms of telescope time and need to rely on a secure
physical link between the source measured and the molecular cloud).
The \textit{statistical\/} error associated with this technique is
linked to $N_\mathrm{fg}$, the number of foreground stars observed in
projection to the cloud.  This quantity, for nearby clouds, increases
linearly with the cloud distance $d$, and therefore the relative
statistical error on the estimated distance goes as $1/\sqrt{d}$.

\subsection{Column density probability distribution}
\label{sec:column-dens-prob}

\begin{figure*}[!tbp]
  \begin{center}
    \includegraphics[width=\hsize]{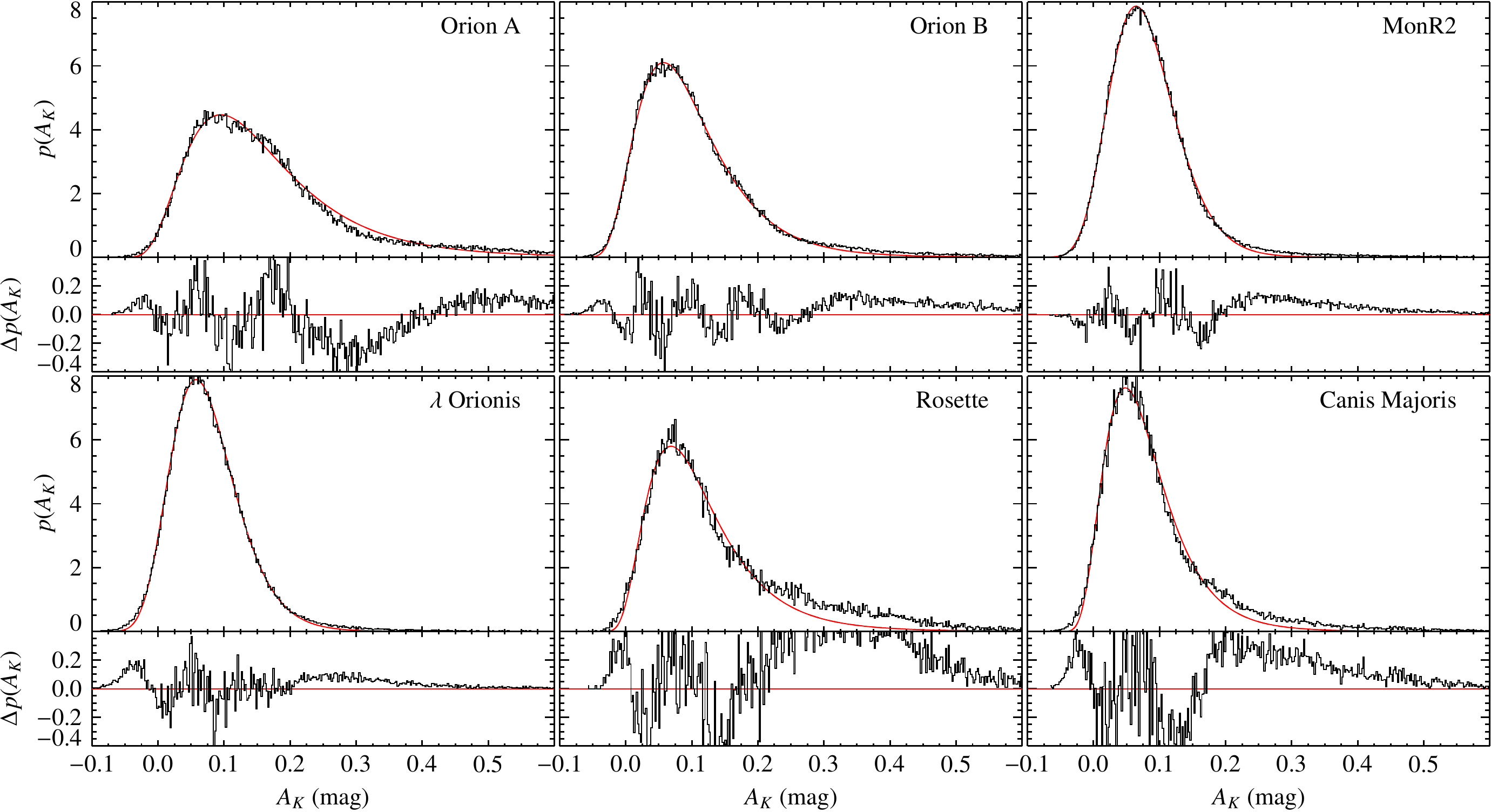}
    \caption{The probability distributions of pixel extinctions for
      the six cloud complexes.  In each plot, the red, solid curve
      represents the best-fit with a log-normal distribution.  Lower
      panels show the residuals with respect to the best-fit.}
    \label{fig:11}
  \end{center}
\end{figure*}

\begin{table}[b!]
  \centering
  \caption{The best-fit parameters of the four Gaussian functions used
    to fit the column density probability distribution shown in
    Figs.~\ref{fig:11} (see Eq.~\eqref{eq:6} for the
    meaning of the various quantities).}
  \label{tab:2}
  \tiny
  \begin{tabular}{lccc}
    Cloud & Offset $A_0$ & Scale $A_1$ & Dispersion $\sigma$ \\
    \hline
    Orion A & $-0.059$ & $0.193$ & $0.491$ \\
    Orion B & $-0.060$ & $0.145$ & $0.482$ \\
    Mon R2  & $-0.163$ & $0.238$ & $0.214$ \\
    $\lambda$ Orionis  & $-0.053$ & $0.121$ & $0.378$ \\
    Rosette & $+0.027$ & $0.079$ & $1.167$ \\
    Canis Major & $+0.013$ & $0.053$ & $0.978$ 
  \end{tabular}
\end{table}

Many theoretical studies have suggested that the turbulent supersonic
motions that are believed to characterize the molecular clouds on
large scales induce a log-normal probability distribution for the
volume density \citep[e.g.][]{1994ApJ...423..681V,
  1997MNRAS.288..145P, 1998PhRvE..58.4501P, 1998ApJ...504..835S}.
Although this result strictly applies for the \textit{volume density},
under certain assumptions, verified in relatively ``thin'' molecular
clouds, the probability distribution for the \textit{column density},
i.e.\ the volume density integrated along the line of sight, is also
expected to follow a log-normal distribution
\citep{2001ApJ...557..727V}.  Additionally,
\citet{2010MNRAS.408.1089T} recently have shown that log-normal
distributions are also expected under completely different physical
conditions (also plausible for molecular clouds), such as radially
stratified density distributions dominated by gravity and thermal
pressure, or by a gravitationally-driven ambipolar diffusion.

The log-normality of the column densities is verified with good
approximation at low-column densities in many of the clouds that we
have studied in the past and by similar results obtained by
\citet{2009A&A...508L..35K}, \citet{2009ApJ...692...91G}, and
\citet{2010MNRAS.406.1350F}.  As shown by the plots in
Fig.~\ref{fig:11}, the region studied here confirms this general
trend.  The probability distributions of column densities for the
various clouds were fitted with a log-normal distributions of the
form\footnote{Note that the functional form used here differs, in the
  definition of $\sigma$, with respect to the form used in the
  previous papers.}
\begin{equation}
  \label{eq:6}
  h(A_K) = \frac{a}{A_K - A_0} 
  \exp\left[- \frac{\bigl(\ln (A_K - A_0) - \ln A_1 \bigr)^2}%
    {2 \sigma^2} \right] \; .
\end{equation}
For some of the clouds, such as Orion B, $\lambda$ Orionis, and
Mon~R2, the fits appear to be better than for other ones, such as
Orion~A or Rosette.  However, in all cases residuals are well above
the expected levels\footnote{The theoretical error follows a Poisson
  distribution, and is therefore different for each cloud and each
  bin.  In the range displayed in Fig.~\ref{fig:11}, the median error
  is approximately $0.1 \mbox{ mag}$, but since different bins are
  expected to be uncorrelated, the systematic offsets shown by the
  various clouds for $A_K > 0.2 \mbox{ mag}$ are highly significant.}
and show systematic and structured deviations even at low column
densities.  Additionally, all clouds show a positive residual at the
higher column densities, approximately for $A_K > 0.2 \mbox{ mag}$.
The significance of these results and the goodness of the fits need to
be further investigated.

One perhaps surprising feature of Fig.~\ref{fig:11} is the presence of
a significant number of column density estimates with negative values.
This could be either due to a zero-point offset in the control field
or to uncertainties in the column density measurements, which
naturally broadens the intrinsic distribution and possibly adds a
fraction of negative measurements.  Note also that the amount of
negative pixels observed is compatible with the typical error on our
extinction maps, which is of the order of $0.03 \mbox{ mag}$.

\subsection{Small-scale inhomogeneities}
\label{sec:small-scale-inhom}

\begin{figure}[!tbp]
  \begin{center}
    \includegraphics[width=\hsize]{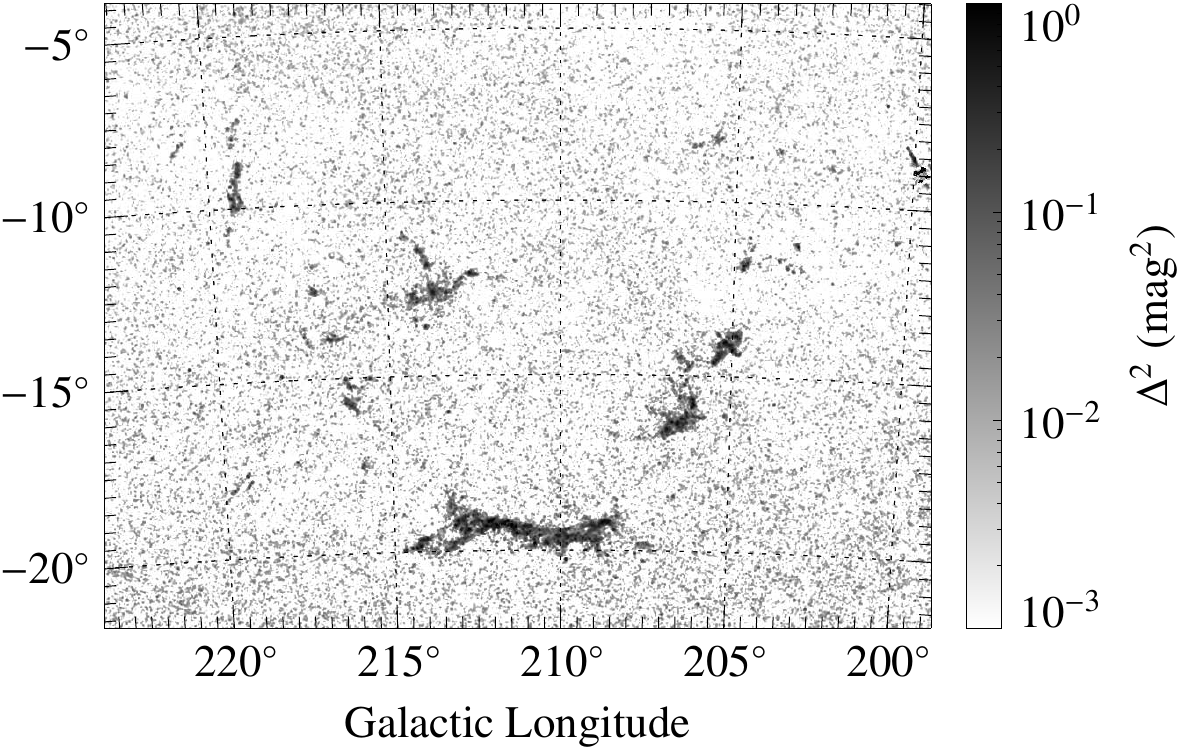}
    \caption{The $\Delta^2$ map of Eq.~\eqref{eq:7} on the same region
      shown in Fig.~\ref{fig:4}.  Note the significant increase
      observed in $\Delta^2$ close to the central parts of this
      cloud.}
    \label{fig:12}
  \end{center}
\end{figure}

\begin{figure}[!tbp]
  \begin{center}
    \includegraphics[width=\hsize]{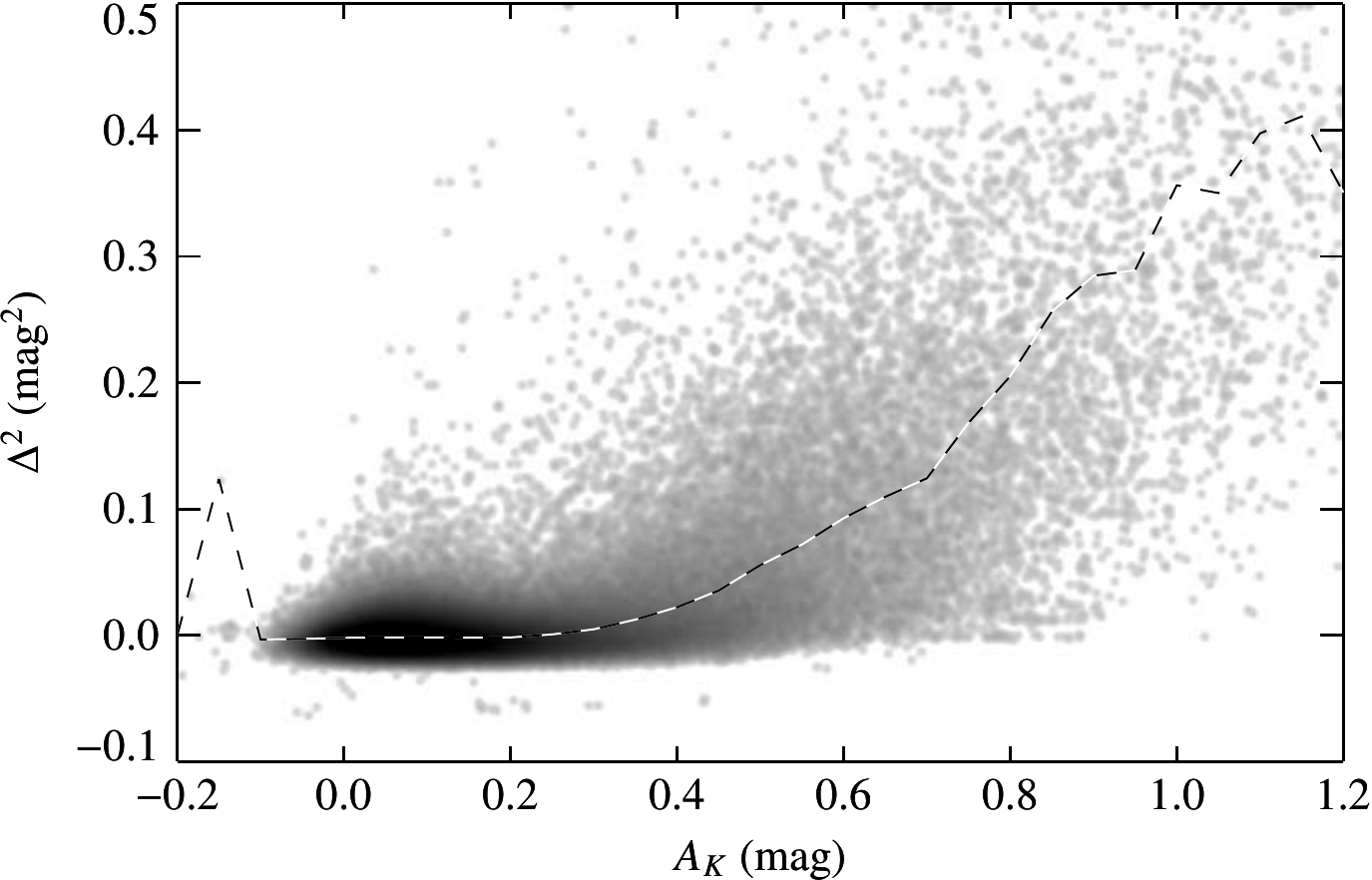}
    \caption{The distribution of the $\Delta^2$ map as a function of
      the local extinction $A_K$ for the map in Fig.~\ref{fig:3}, in
      logarithmic grey scale.  The dashed line shows the average
      values of $\Delta^2$ in bins of $0.05 \mbox{ mag}$ in $A_K$.
      Note the rapid increase of $\Delta^2$ for $A_K > 0.7 \mbox{
        mag}$.  As a comparison, the average variance
      $\mathrm{Var}\bigl( \hat A^{(n)}_K \bigr)$ on the estimate of
      $\hat A_K$ from a single star is approximately $0.033 \mbox{
        mag}^2$.}
    \label{fig:13}
  \end{center}
\end{figure}

\citet{1994ApJ...429..694L} first recognized that the local dispersion
of extinction measurements increases with the column density.  In
other words, within a single ``pixel element,'' the scatter of the
individual stellar column density estimates is proportional to the
average local column density estimate.  This results implies the
presence of substructures on scales smaller than the resolution of the
extinction maps, and shows that theses substructures are more evident
in regions with high column density.  Substructures could be due
either to unresolved gradients or to random fluctuations induced by
turbulence \citep[see][]{1999ApJ...512..250L}

The presence of undetected inhomogeneities is important for two
reasons: (i) they might contain signatures of turbulent motions
\citep[see, e.g.][]{1994ApJ...429..645M, 1997ApJ...474..730P}, and
(ii) they are bound to bias the extinction measurements towards lower
extinctions in high-column density regions (and, especially, in the
very dense cores; see \citealp{2009A&A...493..735L}).

In the previous papers of this series we have considered a quantity
that traces well the inhomogeneities:
\begin{equation}
  \label{eq:7}
  \Delta^2(\vec\theta) \equiv \hat\sigma^2_{\hat A_K}(\vec\theta) +
  \sigma^2_{\hat A_K}(\vec\theta) - 
  \bigl\langle \mathrm{Var}\bigl( \hat A_K^{(n)} \bigr) \bigr\rangle
  (\vec \theta) \; .
\end{equation}
The $\Delta^2$ map is defined in terms of the observed variance of
column density estimates,
\begin{equation}
  \label{eq:8}
  \hat\sigma^2_{\hat A_K}(\vec\theta) \equiv \frac{\sum_{n=1}^N W^{(n)} 
    \bigl[ \hat A_K^{(n)} - \hat A_K(\vec\theta)
    \bigr]^2}{\sum_{n=1}^N W^{(n)}} \; ,
\end{equation}
the average expected scatter due to the photometric errors and the
intrinsic dispersion in the colors of the stars
\begin{equation}
  \label{eq:9}
  \sigma^2_{\hat A_K}(\vec \theta) \equiv \frac{\sum_{n=1}^N \bigl[
    W^{(n)} (\vec\theta) \bigr]^2 \mathrm{Var} \bigl( \hat A_K^{(n)}
    \bigr)}{\Bigl[ \sum_{n=1}^N W^{(n)}(\vec\theta) \Bigr]^2} \; ,
\end{equation}
and of the weighted average expected variance for the column density
measurements around $\vec\theta$
\begin{equation}
  \label{eq:10}
  \bigl\langle \mathrm{Var}\bigl( \hat A_K^{(n)} \bigr) \bigr\rangle
  (\vec \theta) \equiv \frac{\sum_n W^{(n)}
    \mathrm{Var}\bigl( \hat A_K^{(n)} \bigr)}{\sum_n W^{(n)}} \; .
\end{equation}
As shown in Paper~II, the combination of observables that enters the
definition \eqref{eq:7} ensures that the expected value for
$\Delta^2(\vec\theta)$ is a weighted average of the \textit{square\/}
of local inhomogeneities:
\begin{equation}
  \label{eq:11}
  \bigl\langle \Delta^2(\vec\theta) \bigr\rangle = \frac{\sum_n
    W^{(n)} \bigl[ A_K\bigl( \vec\theta^{(n)} \bigr) - \tilde A_K
    \bigr]^2}{\sum_n W^{(n)}} \; .
\end{equation}
The term inside brackets in the numerator of this equation represents
the local scatter of the column density at $\vec\theta^{(n)}$ with
respect to the weighted average column density in the patch of the sky
considered $\tilde A_K \equiv \sum_n W^{(n)} A_K\bigl( \vec\theta^{(n)}
\bigr) \bigm/ \sum_n W^{(n)}$.  Note also that this definition for
$\tilde A_K$ implies that the average of local inhomogeneities
vanishes, as expected [cf.\ Eq.~\eqref{eq:11}]:
\begin{equation}
  \label{eq:12}
  \frac{\sum_n W^{(n)} \bigl[ A_K\bigl( \vec\theta^{(n)} \bigr) -
    \tilde A_K \bigr]}{\sum_n W^{(n)}} = 0 \; .
\end{equation}

Similar to the other papers of this series, we evaluated the
$\Delta^2$ map for the whole field and identified regions with large
small-scale inhomogeneities.  The $\Delta^2$ map for the region
considered in Fig.~\ref{fig:4} is shown in Fig.~\ref{fig:12}.  As
usual, inhomogeneities are mostly present in high column density
regions, while at low extinctions (approximately below $A_K < 0.4
\mbox{ mag}$) substructures are either on scales large enough to be
detected at our resolution ($2.5 \mbox{ arcmin}$), or are negligible.

The average $\Delta^2$ as a function of the local extinction, $A_K$,
for the map in Fig.~\ref{fig:6} is presented in Fig.~\ref{fig:13}.  A
comparison of the dashed line, representing the average value of
$\Delta^2$ in bins of $0.05 \mbox{ mag}$ in $A_K$, with the average
variance $\mathrm{Var}\bigl( \hat A^{(n)}_K \bigr)$ on the estimate of
$\hat A_K$ from a single star, which is approximately $0.033 \mbox{
  mag}^2$, shows that unresolved substructures start to be the
prevalent source of errors in extinction maps for $A_K > 0.5 \mbox{
  mag}$.  As discussed in \cite{2009A&A...493..735L}, this column
density value gives also an approximate upper limit for the robustness
of \textsc{Nice} and \textsc{Nicer} extinction studies, since both
these methods are based on the implicit assumption that the extinction
is uniform within a resolution element (in our case, within the region
where the window function $W(\vec\theta)$ is significantly different
from zero).  In contrast, the \textsc{Nicest} algorithm is explicitly
designed to cope with unresolved substructures, and therefore it
should be used in the region considered when $A_K > 0.5 \mbox{ mag}$.
We stress that the use of \textsc{Nicest} does not remove the
unresolved inhomogeneities (and thus does not make the $\Delta^2$ map
flat), but rather it makes sure that the estimate of $A_K$ is unbiased
even if $\Delta^2$ is non-vanishing.

\section{Mass estimate}
\label{sec:mass-estimate}

\begin{table}[b]
  \centering
  \def\0{\phantom{0}}
  \begin{tabular}{lcccc}
    Cloud    & Distance &
    \multicolumn{3}{c}{Mass ($\mbox{M}_\odot$)} \\
    \cline{3-5}
             &          & Total & $A_K > 0.1$ & $A_K > 0.2$ \\
    \hline
    Orion A & $\0371 \mbox{ pc}$ & $\075\,700$
    & $\066\,400$ & $\045\,100$ \\
    Orion B & $\0398 \mbox{ pc}$ & $\095\,100$
    & $\068\,300$ & $\036\,100$ \\
    Mon R2  & $\0905 \mbox{ pc}$ & $ 392\,000$
    & $ 222\,000$ & $\073\,300$ \\
$\lambda$ Ori&$\0445 \mbox{ pc}$ & $ 102\,000$
    & $\047\,400$ & $\011\,500$ \\
    Rosette & $ 1330 \mbox{ pc}$ & $ 233\,000$
    & $ 193\,000$ & $ 137\,000$ \\
Canis Major & $ 1150 \mbox{ pc}$ & $ 205\,000$
    & $ 137\,000$ & $\076\,900$ \\
  \end{tabular}
  \caption{The masses of all complexes studied in this paper.  The
    table reports the assumed distance (cf.\ Table~\ref{tab:1}),
    followed by the estimated masses over the whole area, and over areas
    within the $A_K > 0.1 \mbox{ mag}$ and the $A_K > 0.2 \mbox{ mag}$
    contours.}
  \label{tab:3}
\end{table}

Masses of the clouds were evaluated using the standard relation
\begin{equation}
  \label{eq:13}
  M = d^2 \mu \beta_K \int_\Omega A_K(\vec\theta) \, \diff^2 \theta \; ,
\end{equation}
where $d$ is the cloud distance, $\mu$ is the mean molecular weight
corrected for the helium abundance, and $\beta_K \simeq 1.67 \times
10^{22} \mbox{ cm}^{-2} \mbox{ mag}^{-1}$ is the ratio $\bigl[
N(\mathrm{H\textsc{i}}) + 2N(\mathrm{H}_2) \bigr] / A_K$
(\citealp{1979ARA&A..17...73S}; see also \citealp{1955ApJ...121..559L,
  1978ApJ...224..132B}).  Assuming a standard cloud composition
($63\%$ hydrogen, $36\%$ helium, and $1\%$ dust), we find $\mu =
1.37$.  The integral above was evaluated either over the whole area of
each cloud, as defined in Eqs.~\eqref{eq:5}, or inside contours above
a given extinction threshold.  This latter option allowed us to avoid
using the ``total'' mass of a cloud, which is ill defined and strongly
depends on the contours chosen, in favor of the mass of a cloud inside
specified extinction contours.

We stress that the masses indicated for the various clouds refers to
the boundaries taken in this paper.  In particular, if we increase the
southern boundary of MonR2 from $b > -17^\circ$ to $b > -15^\circ$,
thus avoiding the filament south of NGC~2149, then the mass estimates
for MonR2 decrease to $311\,000 \mbox{ M}_\odot$ (total), $166\,000
\mbox{ M}_\odot$ ($A_K > 0.1 \mbox{ mag}$), and $58\,600 \mbox{
  M}_\odot$ ($A_K > 0.2 \mbox{ mag}$).  Correspondingly, if this area
encloses Orion~A, then the masses of this complex increase to $86\,000
\mbox{ M}_\odot$ (total), $75\,800 \mbox{ M}_\odot$ ($A_K > 0.1 \mbox{
  mag}$), and $47\,600 \mbox{ M}_\odot$ ($A_K > 0.2 \mbox{ mag}$).

Uncertainties on the cloud masses are difficult to evaluate, because
they are essentially dominated by systematic errors.  Note that the
errors on the mass due to statistical noise of the extinction maps are
negligible.  Instead, the major sources of errors are the uncertainty
on the cloud distances (which, because of the dependence of the mass
on the square of the distance, can be relevant), and possible
systematic errors on the zero-point of the extinction maps (typically
due to a residual extinction present in the control field).

\begin{figure}[!tbp]
  \begin{center}
    \includegraphics[width=\hsize]{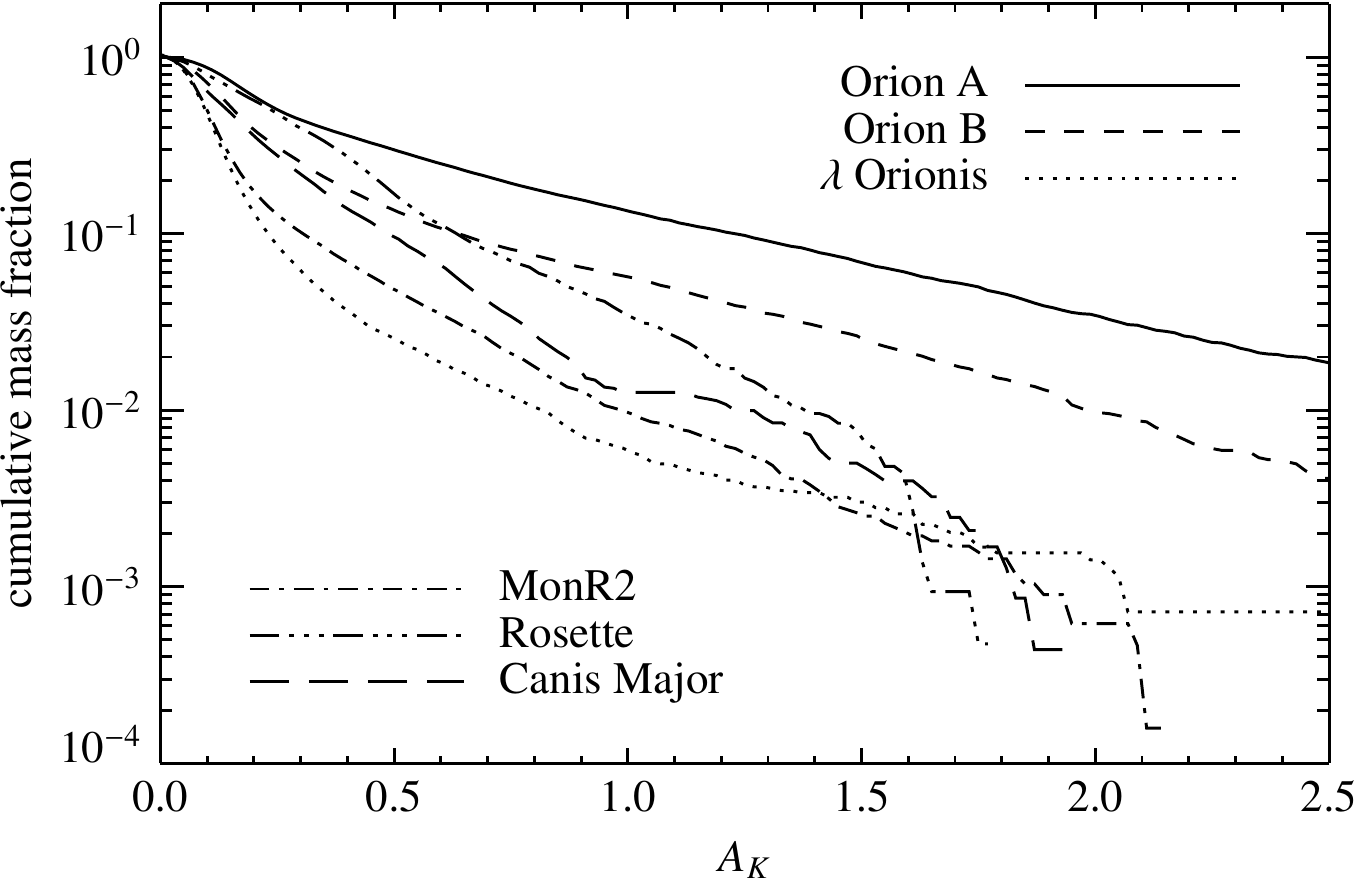}
    \caption{The cumulative mass enclosed in iso-extinction contours
      for the various clouds.  All plots have been obtained from the
      extinction map shown in Fig.~\ref{fig:6}, and have thus the same
      resolution angular limit ($\mathit{FWHM} = 3 \mbox{ arcmin}$).
    \label{fig:14}}
  \end{center}
\end{figure}

\begin{figure}[!tbp]
  \begin{center}
    \includegraphics[width=\hsize]{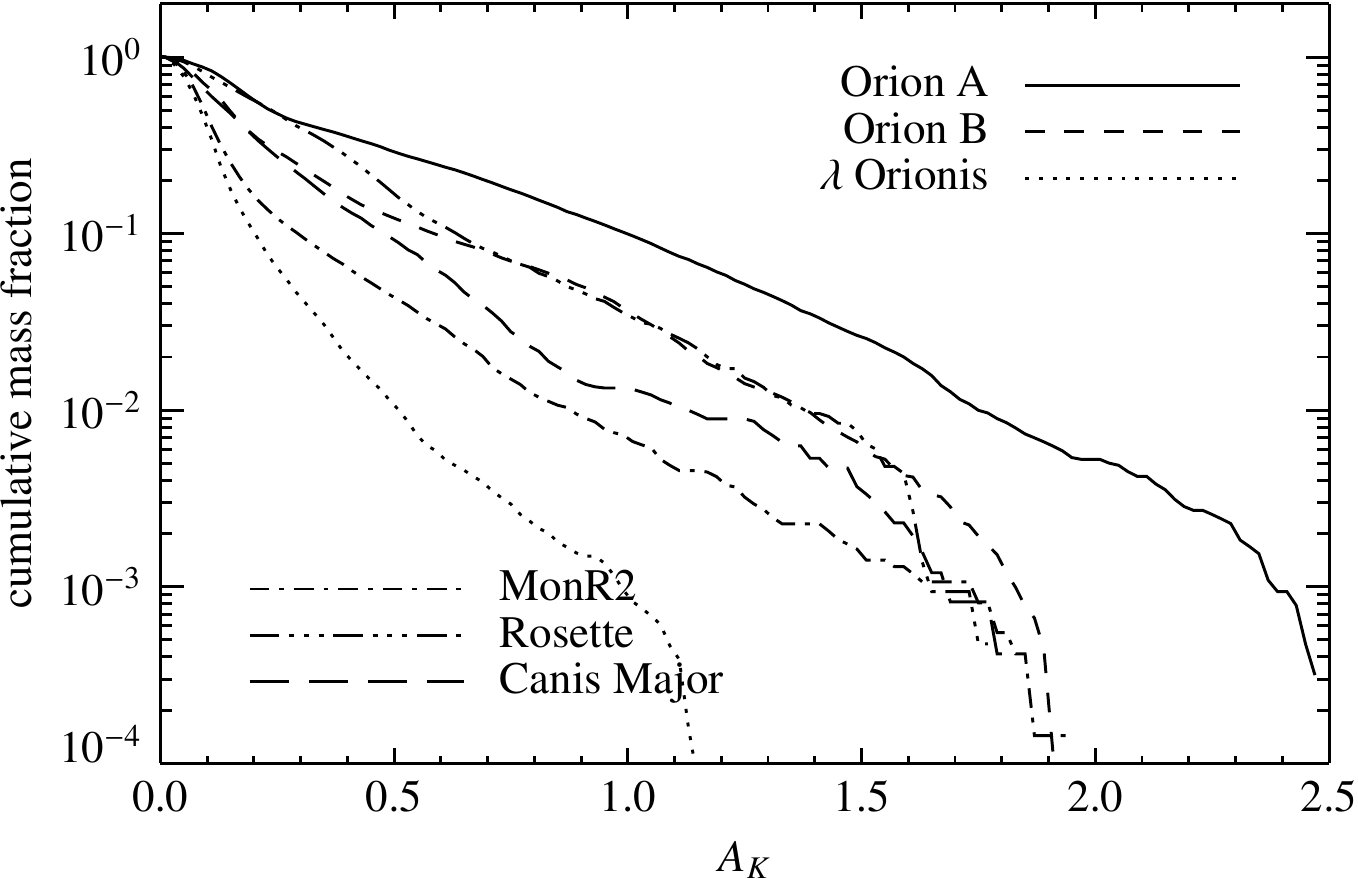}
    \caption{Same as Fig.~\ref{fig:14}, but the extinction maps for
      the various clouds were degraded to the same physical
      resolution $\mathit{FWHM} \simeq 1.13 \mbox{ pc}$.}
    \label{fig:15}
  \end{center}
\end{figure}

As in the other papers in this series, it is interesting to
investigate the cumulative mass as a function of the extinction
threshold.  This plot, shown in Fig.~\ref{fig:14}, provides a simple
measure of the structure of the molecular clouds, and of the relative
importance of low- and high-density regions within each cloud.
Additionally, as shown by \citet{2010ApJ...724..687L}, the cloud mass
at high densities (in particular, for $A_K > 0.8 \mbox{ mag}$) appears
to correlate with the overall star formation rate of the cloud.

Figure~\ref{fig:15} shows the same plot of Fig.~\ref{fig:14}, but
using this time the same physical resolution for all clouds.  That was
accomplished by degrading the extinction map of the nearby clouds to
match the physical resolution of the most distant cloud, the Rosette.
This procedure makes even more evident the difference between the
various complexes; interestingly, the smoothing applied also
strengthens differences between clouds at the same distance, such as
Orion~A and $\lambda$ Orionis.  This last point highlights differences
in the density structure present in the clouds studied here.  In
particular, clouds that present a plateau of relatively high values of
extinction are not affected too much by the smoothing applied in
Fig.~\ref{fig:15}; on the contrary, clouds such as $\lambda$ Orionis
that have dispersed material will have their mass dispersed over an
even larger area, and will therefore present sharp decreases in their
cumulative mass function for relatively small values of $A_K$.

\section{Conclusions}
\label{sec:conclusions}

The following items summarize the main results presented in this
paper:
\begin{itemize}
\item We measured the extinction over an area of $\sim 2\,200$ square
  degrees that encompasses the Orion, the Monoceros R2, the Rosette,
  and the Canis Major molecular clouds.  The extinction map, obtained
  with the \textsc{Nicer} and \textsc{Nicest} algorithms, has a
  resolution of $3 \mbox{ arcmin}$ and an average 1-$\sigma$ detection
  level of $0.3$ visual magnitudes.
\item We measured the reddening laws for the various clouds and showed
  that they agree with the standard \citet{2005ApJ...619..931I}
  reddening law, with the exception of the Mon~R2 region.  We
  interpret the discrepancies observed as a result of contamination
  from a population of young blue stars (likely including members of
  the Orion OB1 association).
\item We estimated the distances of the various clouds by comparing
  the density of foreground stars with the prediction of the
  \citet{2003A&A...409..523R} Galactic model.  All values obtained
  were found to be in very good agreement with independent measurements.
\item We considered the column density probability distributions for
  the clouds and obtained reasonable log-normal fits for all of them.
\item We measured the masses of the clouds and their cumulative mass
  distributions and found differences in the internal structure among
  the clouds.
\end{itemize}

\acknowledgements 

This research has made use of the 2MASS archive, provided by NASA/IPAC
Infrared Science Archive, which is operated by the Jet Propulsion
Laboratory, California Institute of Technology, under contract with
the National Aeronautics and Space Administration.  Additionally, this
research has made use of the SIMBAD database, operated at CDS,
Strasbourg, France.

\bibliographystyle{aa} 
\bibliography{../../dark-refs}

\end{document}